\documentclass[12pt,reqno]{amsart} 
\usepackage{amsthm,amsmath,amssymb,amsfonts,amscd,latexsym,verbatim,xypic}
\usepackage{graphicx}
\usepackage{psfrag}
\newtheorem{theorem}{theorem}[section]
\newtheorem{lemma}[theorem]{Lemma}
\newtheorem{proposition}[theorem]{Proposition}
\newtheorem{corollary}[theorem]{Corollary}
\newtheorem{remark}[theorem]{Remark}
\newtheorem{definition}[theorem]{Definition}
\newcommand{\be} {\begin{equation}}
\newcommand{\ee} {\end{equation}}
\newcommand{\bea} {\begin{eqnarray}}
\newcommand{\eea} {\end{eqnarray}}
\newcommand{\lp}  {\left(}
\newcommand{\rp}  {\right)}
\newcommand{\eqdef} {\stackrel{\rm def}{=}}
\def\bbbone{{\mathchoice {\rm 1\mskip-4mu l} {\rm 1\mskip-4mu l}
{\rm 1\mskip-4.5mu l} {\rm 1\mskip-5mu l}}}
\newcommand{\RR}{\mathbb{R}} 
\newcommand{\ZZ}{\mathbb{Z}}
\newcommand{\NN}{\mathbb{N}}
\newcommand{\CC}{\mathbb{C}}
\newcommand{\KK}{\mathbb{K}}
\newcommand{\cO}{\mathcal{O}}

\newcommand{\cLL}{\mathcal{L}}

\newcommand{\cA}{\mathcal{A}}
\newcommand{\cS}{\mathcal{S}}
\newcommand{\cB}{\mathcal{B}}
\newcommand{\cW}{\mathcal{W}}
\newcommand{\cE}{\mathcal{E}}
\newcommand{\cK}{\mathcal{K}}
\newcommand{\mF}{\mathfrak{m}}
\newcommand{\ep}{\epsilon} 
\newcommand{\om}{\omega}

\newcommand{\ph}{\phi}
\newcommand{\ka}{\kappa}
\newcommand{\vka}{\varkappa}
\newcommand{\de}{\delta}
\newcommand{\De}{\Delta}
\newcommand{\ta}{\tau}
\newcommand{\et}{\eta}
\newcommand{\ga}{\gamma} 
\newcommand{\ze}{\zeta} 
\newcommand{\al}{\alpha}
\newcommand{\la}{\lambda}
\newcommand{\La}{\Lambda}
\newcommand{\Ga}{\Gamma}
\newcommand{\Si}{\Sigma} 
\newcommand{\Ups}{\Upsilon}
\newcommand{\oDe}{\stackrel{\circ}{\De}}
\newcommand{\oX}{\stackrel{\circ}{X}}
\newcommand{\gb}{\bar{g}}
\newcommand{\gbs}{\bar{g}_\ast}
\newcommand{\omz}{\omega_0}
\newcommand{\BSi}{\bar{\Sigma}} 
\begin{document} 
\title[A Complete Renormalization Group
Trajectory]{A Complete Renormalization Group Trajectory
Between Two Fixed Points}
\author[Abdesselam]
{Abdelmalek Abdesselam}
\maketitle

\parbox{12cm}{\small 
{\sc Abstract.} 
We give a rigorous nonperturbative construction
of a massless discrete trajectory for Wilson's exact
renormalization group. The model is a three
dimensional Euclidean field theory with a modified
free propagator. The trajectory realizes the
mean field to critical crossover
from the ultraviolet Gaussian fixed point to an analog
recently constructed by Brydges, Mitter and Scoppola
of the Wilson-Fisher nontrivial fixed point.}

\vspace{5mm} 

\mbox{\small AMS subject classification (2000): 81T08; 81T16; 81T17; 
82B27; 37D10}

\medskip 

\parbox{12cm} 


\section{Introduction}

In recent years, the mathematical community
has shown an increasing interest for
the important but difficult topic
of quantum field theory~\cite{Deligneetal}.
The most comprehensive and insightful, albeit largely conjectural,
mathematical framework to address this subject
is Wilson's renormalization group:
a grand dynamical system in the space of all imaginable
observation scale dependent effective field theories~\cite{WK,WKondo}.
As emphasized by Wilson himself~\cite{WKondo},
his approach can be construed as a
mathematical theory of scaling symmetry which has yet
to be fully unveiled. It generalizes
in a very deep way ordinary calculus which matured in the hands
of 19th century mathematicians and gave the first rigorous meaning
to the notion of `continuum'.
Formulating precise conjectures about the {\em phase portrait}
of the RG dynamical system and proving them, an endeavor
one could perhaps call the `Wilson Program'~\cite{MitterOW},
is one of the greatest challenges in mathematical analysis
and probability theory,
and will likely remain so for years to come.

When studying a phase portrait, the first features to examine
are {\em fixed points}, which here mean scale invariant theories.
If $D$ is the dimension of space, one expects that for $D\ge 4$ there
are only two fixed points:
the high temperature one and the massless Gaussian one.
As one lowers the dimension to the range $3\le D<4$, only one new
fixed point should appear: the Wilson-Fisher fixed point~\cite{WF}.
Its existence as well as the construction of its local
stable manifold, in the hierarchical approximation, was first
rigorously established in~\cite{BleherS};
see also~\cite{ColletE,ColletEbook,GKfix}.
The uniqueness, in the local potential approximation, was shown
in~\cite{Lima}. As one continues lowering the dimension
to the range $2<D<3$, past every threshold
$D_n=2+\frac{2}{n-1}$, $n=3,4,\ldots$, a new fixed point
appears corresponding to an $n$-well potential, as was proved in the
local potential approximation by Felder~\cite{Felder}.
For $D=2$, the situation becomes extremely complicated:
even a conjectural classification of fixed points
corresponding to conformal field theories is not yet complete.
Nevertheless, there have been
tremendous advances in this area;
see e.g.~\cite{DiFranc} and Gaw\c{e}dzki's lectures
in~\cite{Deligneetal} for an introduction.

The next stage in the investigation concerns the various {\em local
invariant manifolds} around these fixed points and the associated
{\em critical exponents}. The first such rigorous result,
for the Gaussian fixed point, is the work of Bleher and
Sinai~\cite{BSGauss}.
For further developments, with emphasis on these dynamical systems
aspects,
see for instance~\cite{ColletE,GKfix,GKcorrel,KochW,Pereira,Wiecz,Domingos}.

Then, in the third stage, one would like
to know more {\em global} features like how all these
local invariant manifolds meet to form separatrices
between domains exhibiting qualitatively different behaviours.
This question pertains to the active field of
the renormalization group theory of crossover phenomena
(see e.g.~\cite{OConnor,Meurice} for recent reviews). Our work
falls within this third class of problems.
The control of a massless RG trajectory between fixed points
announced in~\cite{A} and for which details are provided
here is our contribution
to the grand scheme of the Wilson Program.
Note that there is extensive physics literature, following
the seminal work of Zamolodchikov~\cite{Zamo1,Zamo2},
on such massless RG flows
in particular
in two dimensions, see e.g.~\cite{ZamZam,Delfinoetal,DoreyDT,Dunning} and references therein.
However, nonperturbative results substantiated by rigorous
mathematical estimates
are scarce. To borrow the terminology of the French 
school of constructive field theory, this is the `probl\`eme
de la soudure' or the welding problem.
One has to control the junction between the ultraviolet
and the infrared regimes. For instance, for the two dimensional
Gross-Neveu model, the UV regime has been given a rigorous
mathematical treatment a long time ago~\cite{GKGross,FMRSGross}.
Likewise, the IR regime with spontaneous mass
generation for a {\em UV-cutoff} theory is also under
control~\cite{KMR}; see also~\cite{Kopper} for a similar result
on the sigma model.
However, the junction, although probably not out of reach
of present methods, has proved to be more technically
demanding than expected~\cite{KMRpersonal}. Note that
the model we consider here is simpler in that regard. It does
not involve a drastic change of scenery, for instance, from
a purely Fermionic theory at the ultraviolet end, to a Bosonic one
at the infrared end.

\medskip
Given a small positive bifurcation parameter $\ep$,
we consider a three dimensional $\phi^4$ theory
with a modified propagator: the $(\Phi^4)_{3,\ep}$ model of~\cite{BMS},
which was also studied in the hierarchical approximation in~\cite{GKfix}.
Namely, we consider functional integrals of the form
\be
\int d\mu_{\tilde{C}}(\ph)\ldots e^{-V(\ph)}
\label{functinteg}
\ee
where $d\mu_{\tilde{C}}$ is the Gaussian measure with covariance
$\tilde{C}\eqdef(-\De)^{-\lp\frac{3+\ep}{4}\rp}$
and Wick ordered interaction potential
\be
V(\ph)\eqdef\int_{\RR^3} d^3 x\
\left\{
g :\ph(x)^4:_{\tilde{C}}+ \mu :\ph(x)^2:_{\tilde{C}}
\right\}\ .
\label{potentV}
\ee
Over the last two decades, Brydges and his collaborators
have devised a general mathematical framework,
going beyond the hierarchical and local potential approximations,
in order to give a rigorous nonperturbative
meaning to the renormalization group
dynamical system~\cite{BY,BDH,BDHfix,BMS}. The approach,
actually involving no approximation whatsoever, is 
in the spirit of Wilson's exact renormalization
group scheme~\cite{Mitter}.
Our article which can be viewed as a direct continuation of~\cite{BMS}
takes place in this setting.
In very rough terms,
the renormalization group map, rather than flow,
represents the evolution of the integrand $I(\ph)$ of functional integrals
such as (\ref{functinteg})
under convolution and rescaling. The convolution is with respect
to the Gaussian measure corresponding to Fourier modes $p$ of the field
$\ph$ which are restricted to a range of the form
$L^n\le |p|\le L^{n+1}$,
where the integer $L\ge 2$ is the scale ratio for one RG step.
By rescaling, one can keep the integer $n$ constant,
and make the RG transformation autonomous.
The latter acts on the integrand $I(\ph)$ and produces a new one
$I'(\ph)$. However,
the problem with expressing the renormalization group in terms of its
action on $I (\phi)$ is that $I (\phi)$ does not exist in the infinite
volume limit.  It is essential to express $I (\phi)$ in terms of
coordinates that (i) are well defined in the infinite volume limit and
(ii) carry the exact action of the renormalization group in a
tractable form. The key feature that these coordinates have to express
is that $I(\phi)$ is approximately a product of local functionals of
the field and the action of the renormalization group is also
approximately local. The first step towards these coordinates is to
write $I (\phi)$ via the polymer representation:
\be
I(\ph)=\sum_{\{X_i\}} e^{-V(\La\backslash X,\ph)}
\prod_i K(X_i,\ph)
\label{polyrep}
\ee
where $\La$ is the volume cut-off needed to perform the thermodynamic limit,
and $\{X_i\}$ is a collection of disjoint polymers $X_i$ in $\La$.
By polymer we mean
a connected finite union of cubes cut by a fixed $\ZZ^3$ lattice inside
$\RR^3$.
The union of the $X_i$ has been denoted by $X$, and the functional
$V(\La\backslash X,\ph)$ is given by (\ref{potentV}) except that the
integration domain is the complement $\La\backslash X$ instead of $\RR^3$.
Therefore, the functionals $V$ are determined by the two variables
or couplings $g$ and $\mu$. Now the $K$'s are local functionals
of the field, which means that $K(Y,\ph)$ only depends on the restriction
of $\ph$ to the set $Y$. The knowledge of the integrand $I(\ph)$
amounts to that of the couplings $g,\mu$ together with the collection $K$
of all the functionals $K(Y,\ph)$ corresponding to all possible polymers $Y$.
One also needs a splitting $K=Q e^{-V}+R$ of these functionals
where the $Qe^{-V}$ part is given explicitly in terms of $g,\mu$ only.
In sum, the integrand is encoded by a triple $(g,\mu,R)$.
The renormalization group map
in~\cite{BMS} is implemented as a mathematically precise transformation
$(g,\mu,R)\mapsto(g',\mu',R')$.
The evolution for the $:\ph^4:$ coupling $g$ has the form
\be
g' =  L^\ep g-L^{2\ep} a(L,\ep) g^2+\xi_{g}(g,\mu,R)\ .
\label{gevol}
\ee
The evolution
of the mass term or $:\ph^2:$ coupling $\mu$ has the form
\be
\mu'  =  L^{\frac{3+\ep}{2}} \mu+\xi_{\mu}(g,\mu,R)\ .
\label{muevol}
\ee
Finally the collection $R$ of `irrelevant terms',
living in a suitable infinite dimensional space, evolves according to
\be
R'  =  \cLL^{(g,\mu)}(R)+\xi_{R}(g,\mu,R)
\label{Revol}
\ee
where $\cLL^{(g,\mu)}$ is a $(g,\mu)$-dependent contractive linear map
in the $R$ direction.
The $\xi$ remainder terms are higher order small nonlinearities.
An important feature of this formalism is that the polymer representation
(\ref{polyrep}) is not unique. As a result, one has enough freedom
when defining the RG map, in order to secure the contractive property
of the $\cLL^{(g,\mu)}$. This is the so called `extraction step'
which encapsulates the renormalization substractions familiar
in quantum field theory. 
The transformation in~\cite{BMS} also carried an
extra dynamical variable $\mathbf{w}$ with very simple
evolution which is independent of the other variables,
and converging exponentially fast to a fixed point $\mathbf{w}_\ast$.
This was introduced in order to make the RG map autonomous.
Throughout this article however, we take
$\mathbf{w}=\mathbf{w}_\ast$ and incorporate $\mathbf{w}$ in the very
definition of the RG map.
In~\cite{BMS}, it was shown that for small $\ep>0$
there exists an infrared fixed point $(g_\ast,\mu_\ast,R_\ast)$
which is an analog of the Wilson-Fisher fixed point~\cite{WF},
and which is nontrivial, i.e., distinct from the Gaussian
ultraviolet fixed point $(g,\mu,R)=(0,0,0)$.
The local stable manifold of the infrared fixed point was also
constructed.
Note that if one neglects the $\xi$ remainders, one gets
an approximate fixed point $(\gbs,0,0)$
where
\be
\gbs\eqdef\frac{L^\ep-1}{L^{2\ep} a(L,\ep)}=\cO(\ep)
\ .
\ee
\begin{figure}[h]
\psfrag{A}{$\mu$}
\psfrag{B}{${\rm Gaussian}\atop{\rm UV\ f.p.}$}
\psfrag{C}{$R$}
\psfrag{D}{$g$}
\psfrag{E}{${\rm BMS}\atop{\rm IR\ f.p.}$}
\psfrag{F}{${\rm critical\ surface}\atop{\ }$}
\centering
\includegraphics[width=12cm]{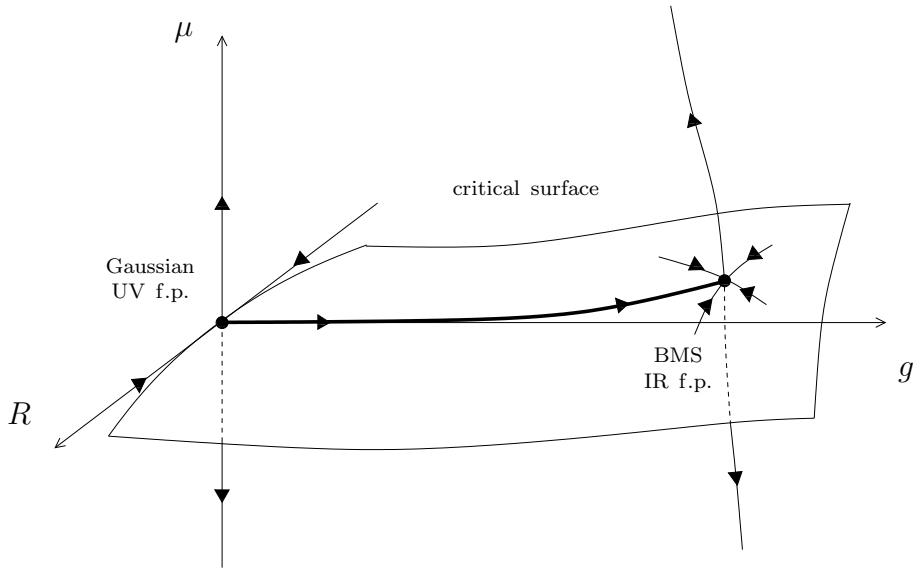}
\caption{The RG dynamical system}
\label{RGpic}
\end{figure}
\noindent
A schematic rendition of the phase portrait of the RG map
considered in~\cite{BMS} is provided by Figure~\ref{RGpic}.
The precise statements of our main results, 
Theorem \ref{mainthm}
and Corollary \ref{fprecovery} below, require a substantial
amount of machinery to be provided in the next sections.
We can nevertheless already give an informal statement.

\noindent{\bf Main result}

{\it
In the regime where $\ep>0$ is small enough, for any
$\omz\in]0,\frac{1}{2}[$, there exists a complete
trajectory $(g_n,\mu_n,R_n)_{n\in\ZZ}$
for the RG map given by Equations (\ref{gevol}),
(\ref{muevol}), and (\ref{Revol}),
such that
$\lim\limits_{n\rightarrow -\infty}
(g_n,\mu_n,R_n)=(0,0,0)$
the Gaussian ultraviolet fixed point, and
$\lim\limits_{n\rightarrow +\infty}
(g_n,\mu_n,R_n)=(g_\ast,\mu_\ast,R_\ast)$
the BMS nontrivial infrared fixed point,
and
determined by the `initial condition' at unit scale
\be
g_0=\omz\gbs\ .
\ee
}
\begin{figure}[h]
\psfrag{a}{${\rm Gaussian}\atop{\rm UV\ f.p.}$}
\psfrag{b}{$\cdots P_{-2}$}
\psfrag{c}{$P_{-1}$}
\psfrag{d}{$P_0$}
\psfrag{e}{$P_1$}
\psfrag{f}{$P_2\cdots$}
\psfrag{g}{${\rm BMS}\atop{\rm IR\ f.p.}$}
\psfrag{h}{$g$}
\psfrag{i}{$\gbs$}
\psfrag{j}{$\mu,R$}
\centering
\includegraphics[width=12cm]{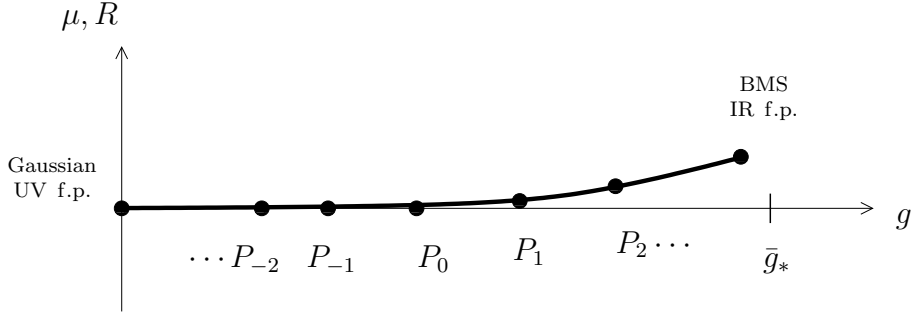}
\caption{The discrete trajectories}
\label{disctraj}
\end{figure}
\noindent
See Figure~\ref{disctraj} for a sketch of such
discrete RG orbits $P_n=(g_n,\mu_n,R_n)$, $n\in\ZZ$,
which are parametrized by the projection of $P_0$ on the $g$ axis.
To the best of our knowledge, the only previous similar
result is the construction of the massless
connecting heteroclinic orbit
going from a UV nontrivial fixed point to
the Gaussian IR fixed point for a modified
Gross-Neveu model in~\cite{GKnonren} (see also~\cite{dCetal,Paolo}
for related work in the massive case).
Our work which essentially amounts to the construction
of a nontrivial massless three dimensional
Euclidean field theory in the continuum,
is probably the first such result in the Bosonic case.
This field theory is superrenormalizable in the ultraviolet sector
but only barely. Namely, one needs to renormalize divergent
Feynman diagrams only up to a finite order in perturbation theory;
however this order goes to infinity when the parameter $\ep$
goes to zero. As shown in~\cite[Section 1.1]{BMS}, a proof for
the difficult axiom of
Osterwalder-Schrader positivity seems feasible on this model, which
makes it interesting from the point of view of traditional
constructive field theory~\cite{GlimmJ}.
Due to the lack of a nonperturbative definition of dimensional regularization,
this model is the best available for the mathematically rigorous
study of the Wilson-Fisher fixed point~\cite{WF} which is believed to govern
the infrared behavior of the tridimensional Ising model (when $\ep=1$). 
On the technical side, as far as the construction of a global RG trajectory
is concerned, one should
note that the situation in~\cite{GKnonren} is facilitated
by the availability of a convergent series representation
in a whole neighborhood of the Gaussian fixed point
which is only possible for a Fermionic theory.
In the present situation, the `trivial' fixed point
around which the analysis takes place is not so trivial
and in fact is highly singular from the point of view of the estimates
we use. This is a manifestation
of the so-called `large field problem' and the need for
the `domination procedure' (see e.g. ~\cite{Rivass}).
The norms needed for the control of $R$ which implement
a measurement of the typical size of the
field $\ph\sim g^{-\frac{1}{4}}$
through a parameter $h$ appearing in the definition
of these norms, create one of the main
difficulties we had to overcome: the `fibered norm problem'.
Namely, the norm for $R$ involves the {\em dynamical variable} $g$.
The approach we used is to construct the trajectory
$s=(g_n,\mu_n,R_n)_{n\in\ZZ}$ via its deviation
$\de s$ with respect to an approximate trajectory
$(\gb_n,0,0)_{n\in\ZZ}$ which solves the RG recursion when
the $\xi$ terms are thrown out.
This is done thanks to a contraction mapping
argument in a big Banach space of sequences $\de s$.
This approach, in the spirit
of Irwin's proof of the stable manifold theorem~\cite{Irwin,Shub},
was suggested to us by D.~C.~Brydges.
We then realized that one can resolve the vicious circle
entailed by the `fibered norm problem' by using the approximate
values $\gb_n$ in the definition of the norms.

In principle,
Wilson's RG picture reduces deep questions
in quantum field theory and statistical mechanics to
a chapter in the theory of bifurcations and dynamical
systems. In practice, it has proved hard to
get away with
the application of a ready-made theorem from the
corresponding literature,
as emphasized in~\cite[p. 70]{ColletE} from the
beginning of the subject and even for the simpler hierarchical
models.
Most of the works on the rigorous renormalization
group use an ad hoc method developed in~\cite{BleherS}.
An innovation was introduced in~\cite{BDHfix}, by
the construction of the stable manifold of the nontrivial
fixed point using an iteration in a space of sequences,
along the lines of Irwin's proof.
The latter method seems more robust and easier to adapt to
our present setting than the more standard Hadamard graph transform
method~\cite{HirschP,Shub}.
Formally, the RG map given by (\ref{gevol}),
(\ref{muevol}), and (\ref{Revol}), with bifurcation parameter $\ep$
corresponds to a {\em transcritical bifurcation},
according to the classification given e.g. in~\cite[p. 177]{Chow}.
The moving nontrivial fixed point goes through the Gaussian 
one as one increases the $\ep$ parameter.
The negative $\ep$ region is forbidden however, since
it would put the nontrivial fixed point in the undefined
$g<0$ region.
Most pertinent to the construction of a connecting heteroclinic
orbit between RG fixed points, in the dynamical systems literature,
is the article~\cite{KopellH}, which is based on Kelley's
center manifold theorem~\cite{Kelley,Carr,Sijbrand}.
However, we have so far been unable to apply these methods
in the present situation.

In the same way~\cite{GKnonren} is based on the
hard analysis estimates of~\cite{GKGross},
our proof is based on Theorem \ref{BMSestimates} below
which summarizes a slight adaptation of the estimates
in~\cite[Section 5]{BMS} built on the techniques
of~\cite{BY,BDH}.
With the exception of the proof of
this theorem which needs a working knowledge of~\cite[Section 5]{BMS},
our article can be
read with only modest prerequisites in functional analysis
as covered e.g. in~\cite{Adams,Berger,Dieudonne}, and in the
theory of Gaussian probability measures in
Hilbert spaces~\cite{Bogachev,Skorohod}.
We provided a completely self-contained
definition of the renormalization group
map $(g,\mu,R)\mapsto(g',\mu',R')$
in Sections \ref{setting}, \ref{functspace}, and \ref{algebra}.
Apart from making the so called extraction step explicit,
this gives us the opportunity to correct some minor
sign and numerical factor errors, but also one serious error,
namely that in~\cite{BMS} the Banach fixed point theorem
was used for a normed space that is not complete.
Fortunately, we
obtained, through discussions with D.~C.~Brydges and P.~K.~Mitter,
an amendment which is provided in Section \ref{functspace}.
It has the advantage that all the estimates in~\cite[Section 5]{BMS}
hold in this new setting without the need for a touch up.
For more efficiency, in the sections defining the RG map,
we adopted a rather terse style of presentation.
We refer the newcomer seeking a proper motivation for this formalism
to~\cite{Mitter} and the introductory sections of~\cite{BY,BDH,BMS}.
Note that these definitions are quite involved and by no means
the first that would come to one's mind. Nevertheless,
they are about the simplest which give a rigorous
nonperturbative meaning to Wilson's exact renormalization group,
and at the same time navigate around the pitfalls of
more na\"{\i}ve approaches. These pitfalls
have been mapped by the pioneering work of
Balaban, Federbush, Feldman, Gallavotti, Gaw\c{e}dzki,
Glimm, Jaffe, Kupiainen, Magnen, Rivasseau, Seiler, S\'en\'eor,
Spencer,
and many others we apologize for not citing~\cite{GlimmJ,Frohlich}.
For more ample introduction to the rigorous renormalization
group than we provide here, the reader from other areas
of mathematics may most profitably
read~\cite{Sokal,Salmhofer,GallavottiRMP,Battle} and Gaw\c{e}dzki's lecture
in~\cite{Deligneetal}
for a first contact. More technical or specialized material
is covered in~\cite{BenfattoG,GlimmJ,Rivass}.


\section{The general setting}
\label{setting}

The ambient space for the field theory we are considering is
Euclidean $\RR^3$.
Given an element $x=(x_1,x_2,x_3)\in\RR^3$
we will use the notation $|x|_\infty\eqdef\max(|x_1|,|x_2|,|x_3|)$
and $|x|_2\eqdef\sqrt{x_1^2+x_2^2+x_3^2}$.
Let $\ep$ be a small nonnegative number,
then with a slight abuse of notation
the kernel of the covariance operator
$\tilde{C}=(-\De)^{-\lp \frac{3+\ep}{4}\rp}$,
which is formally
\be
\tilde{C}(x,y)=\tilde{C}(x-y)=\int_{\RR^3}
\frac{d^3 p}{(2\pi)^3} e^{ip(x-y)}
(p^2)^{-\lp \frac{3+\ep}{4}\rp}\ ,
\ee
is given (see e.g.~\cite[Section II.3.3]{GelfandS}
for a careful derivation)
for noncoinciding points
by the Riesz potential
\be
\tilde{C}(x-y)=\frac{\vka_\ep}{|x-y|_2^{\frac{3-\ep}{2}}}
\ ,
\ee
with
\be
\vka_\ep\eqdef\pi^{-\frac{3}{2}}
\times 2^{-\lp \frac{3+\ep}{2}\rp}\times
\frac{\Ga\lp\frac{3-\ep}{4}\rp}{\Ga\lp\frac{3+\ep}{4}\rp}\ .
\label{varkapdef}
\ee

Let $\varpi:\RR^3\rightarrow\RR$ be a pointwise
nonnegative $C^\infty$ and rotationally invariant function
which vanishes when $|x|_2\ge \frac{1}{2}$ and is
equal to one when $|x|_2\le \frac{1}{4}$.
Let $\tilde{u}\eqdef\varpi\ast\varpi$ be the convolution
of $\varpi$ with itself.
It is nonnegative both in direct and momentum spaces,
and also rotationally invariant.
Since $\varpi(0)>0$, the integral
\[
\int_{\RR^3} d^3 z
\ |z|_2^{-\frac{3}{2}} \varpi(z)
\]
is strictly positive.
We define the function $u_0$ to be the unique
positive multiple of $\varpi$
such that
\be
\int_{\RR^3} d^3 z
\ |z|_2^{-\frac{3}{2}} u_0(z)=\vka_0=(2\pi)^{-\frac{3}{2}}\ .
\ee
The $u_0$ function is fixed once and for all in this article.
Now define
\be
\la_\ep\eqdef\frac{\vka_\ep}
{\int_{\RR^3} d^3 z
\ |z|_2^{-\lp\frac{3+\ep}{2}\rp} u_0(z)}\ ,
\ee
and let $u_\ep(x)=\la_\ep u_0(x)$.
Now we clearly have $\la_\ep\rightarrow 1$ when $\ep\rightarrow 0$
and for $x\neq y$ in $\RR^3$
\be
\int_{0}^{+\infty}
\frac{dl}{l}\ l^{-\lp\frac{3-\ep}{2}\rp}
u_\ep\lp\frac{x-y}{l}\rp
=\frac{\vka_\ep}{|x-y|_2^{\frac{3-\ep}{2}}}
=\tilde{C}(x-y)\ ,
\label{scaledim}
\ee
i.e., the canonically normalized noncutoff covariance.
We now define the scale one UV-cutoff covariance $C$
by
\be
C(x-y)\eqdef\int_{1}^{+\infty}
\frac{dl}{l}\ l^{-\lp\frac{3-\ep}{2}\rp}
u_\ep\lp\frac{x-y}{l}\rp
\ .
\ee

\begin{remark}
\label{uremark}
In~\cite{BMS} the $u_\ep$ is fixed whereas
here it is a variable multiple of a fixed function $u_0$.
Since in the regime where $\ep$ is small the
multiplier $\la_\ep$ can be assumed to be say
between $0.9$ and $1.1$; this has no effect on the
estimates in~\cite{BMS} such as the large field
stability bounds: Equation 2.3, Lemma 5.3 and Lemma 5.4 therein. 
\end{remark}

Let $L\ge 2$ be an integer.
We will also need the fluctuation covariance
\be
\Ga(x-y)\eqdef\int_{1}^{L}
\frac{dl}{l}\ l^{-\lp\frac{3-\ep}{2}\rp}
u_\ep\lp\frac{x-y}{l}\rp\ .
\label{fluctgamdef}
\ee
Note that in Equation (\ref{varkapdef}) the letter `Gamma' denoted the
usual Euler gamma function; however, from now on the notation will be reserved
for the fluctuation covariance (\ref{fluctgamdef}).
The engineering scaling dimension of the field $\ph$
which is denoted by $[\ph]$
is defined by the property $\tilde{C}(l x)=l^{-2[\ph]}\tilde{C}(x)$.
One can read it off Equation (\ref{scaledim}): $[\ph]=\frac{3-\ep}{4}$. 
As in \cite{BMS} we use the notation
\be
C_L(x)\eqdef L^{2[\ph]} C(Lx)
\ee
for scaling of covariances.
We define
$v^{(2)}(x)\eqdef C_L(x)^2-C(x)^2$ and
let
\be
a(L,\ep)\eqdef 36 \int_{\RR^3} d^3 x
\ v^{(2)}(x)\ .
\ee
It is a simple exercise in analysis to
show that, regardless of the precise shape
of the initial cutoff function $u_0$,
one has
\be
\lim\limits_{\ep\rightarrow 0}
a(L,\ep)=a(L,0)=\frac{\log L}{18\pi^2}
\label{asymptot}
\ee
as expected for the second order coefficient
of the beta function of a marginal (at $\ep=0$)
coupling.
As a result, the approximate fixed point
\be
\gbs=\frac{L^\ep-1}{L^{2\ep} a(L,\ep)}
\ee
satisfies
\be
\gbs\sim
18\pi^2 \ep
\label{gbsequiv}
\ee
when $\ep\rightarrow 0$.

Now consider the lattice $\ZZ^3$ inside $\RR^3$.
A {\em unit box} is any closed cube of the form
$[m_1,m_1+1]\times[m_2,m_2+1]\times[m_3,m_3+1]$
with $m=(m_1,m_2,m_3)\in\ZZ^3$.
The set of all unit boxes is denoted by ${\rm Box}_0$.
A nonempty connected subset of $\RR^3$ which is a finite union
of unit boxes is called a {\em polymer}.
The denumerable set of all polymers is denoted by ${\rm Poly}_0$.
We will also need the set ${\rm Poly}_{-1}\eqdef
\{L^{-1}X|X\in{\rm Poly}_0\}$ whose elements are called
$L^{-1}$-{\em polymers}, as well as
 ${\rm Poly}_{+1}\eqdef
\{L.X|X\in{\rm Poly}_0\}$, whose elements are called
$L$-{\em polymers}. Unless otherwise specified, by polymer
we will always mean a unit polymer, i.e., an element of ${\rm Poly}_0$.
For a polymer $X$, we denote $|X|\eqdef{\rm Vol}(X)$
which is also the number of unit boxes in $X$.
We also define its $L$-{\em closure}
$\bar{X}^L$ as the union of all boxes of size $L$ , cut by
the $(L\ZZ)^3$ lattice,
which
contain a unit box in $X$. This is the same as the smallest
$L$-polymer containing $X$, which explains the terminology.
A polymer $X\in {\rm Poly}_0$ with $|X|\le 8$ is called 
a {\em small polymer}. A polymer $X\in {\rm Poly}_0$ with $|X|\le 2$ is 
called an {\em ultrasmall polymer}. A {\em large} polymer simply is
one which is not small.
We finally define the {\em large set regulator} which is a function
$\cA:{\rm Poly}_0\rightarrow\RR_+^\ast$, by
$\cA(X)\eqdef L^{5|X|}$.


\section{Functional spaces}
\label{functspace}

\subsection{Sobolev spaces with gluing conditions}

To each $X\in{\rm Poly}_0$, we associate
a real separable Hilbert space ${\rm Fld}(X)$ where the fields
$\ph:X\rightarrow\RR$ will live.
Given any {\em open} unit box $\oDe$,
with $\De\in{\rm Box}_0$, we consider
the standard Sobolev space $W^{4,2}(\oDe)$
with the norm
\be
||\ph||_{W^{4,2}(\oDe)}\eqdef
\lp
\sum_{|\nu|\le 4}
||\partial^\nu \ph||_{L^2(\oDe)}^2
\rp^{\frac{1}{2}}\ .
\ee
Since obviously $\oDe$ satisfies the so called
strong local Lipschitz condition, by the Sobolev embedding
theorem~\cite[Theorem 4.12]{Adams}
one has a continuous injection
\[
W^{4,2}(\oDe)\hookrightarrow C^2(\De)
\]
where $C^2(\De)$ is the real Banach space
of functions $\ph:\De\rightarrow\RR$
which are of class $C^2$ in the open box $\oDe$ and which
are continuous together with their first and
second derivatives on all of the closed box $\De$.
The norm used on $C^2(\De)$ is the standard one
\be
||\ph||_{C^2(\De)}
\eqdef\sup\limits_{x\in\De}
\max\limits_{|\nu|\le 2}
|\partial^\nu \ph(x)|\ .
\ee
Besides there is a constant $C_{\rm Sobolev}$
independent of the choice of $\De$ in ${\rm Box}_0$, such that
\be
||\ph||_{C^2(\De)}\le
C_{\rm Sobolev}
||\ph||_{W^{4,2}(\oDe)}\ .
\ee

Now define $\widetilde{\rm Fld}(X)$
to be the finite direct sum of the Hilbert spaces
$W^{4,2}(\oDe)$ for $\De$ contained in $X$.
We let ${\rm Fld}(X)$ be the
subspace of $\widetilde{\rm Fld}(X)$
obtained by imposing the following {\em gluing conditions}.
A field $\ph=(\ph_\De)_{\De\subset X}$ belongs
to ${\rm Fld}(X)$ if and only if, for any neighbouring
boxes $\De_1$, $\De_2$ in $X$, the $C^2$ images by the
Sobolev embedding of $\ph_{\De_1}$ and $\ph_{\De_2}$
coincide as well as their first and second derivatives,
on the common boundary component $\De_1\cap\De_2$.
Again by the embedding theorem, this is a closed condition,
and ${\rm Fld}(X)$ is a real Hilbert space
with the norm
\be
||\ph||_{{\rm Fld}(X)}
\eqdef\lp
\sum_{\De\subset X}
\sum_{|\nu|\le 4}
||\partial^\nu \ph_\De||_{L^2(\oDe)}^2
\rp^{\frac{1}{2}}\ .
\ee
Note that any polymer $X$ is the closure of its interior.
Hence, if one lets as before $C^2(X)$ 
be the space of functions $\ph:X\rightarrow\RR$
which are of class $C^2$ in the, possibly disconnected,
open set $\oX$ and which
are continuous together with their first and
second derivatives on the closed connected set $X$;
and if the norm used on $C^2(X)$ is again
the standard one
\be
||\ph||_{C^2(X)}
\eqdef\sup\limits_{x\in X}
\max\limits_{|\nu|\le 2}
|\partial^\nu \ph(x)|\ ;
\ee
then it is not difficult to show that one has an embedding
\be
{\rm Fld}(X)\hookrightarrow C^2(X)
\label{C2Xembed}
\ee
and an inequality
\be
||\ph||_{C^2(X)}\le
C_{\rm Sobolev}
||\ph||_{{\rm Fld}(X)}\ .
\ee
The important thing here is that the constant
is independent of $X$.
We will often regard $\ph$ as a single function on $X$.

\begin{remark}
With this definition the Lemmata~\cite[Lemma 5.1, Lemma 5.2]{BMS}
which are used for pointwise estimation of the fields,
remain valid. The polygonal line arguments needed in
~\cite[Lemma 5.1]{BMS} as well as
~\cite[Lemma 15]{BDH}
on which~\cite[Lemma 5.24]{BMS} rests, are also preserved.
\end{remark}

Now we will also need the notation
\be
||\ph||_{X,1,4}
\eqdef\lp
\sum_{\De\subset X}
\sum_{1\le|\nu|\le 4}
||\partial^\nu \ph_\De||_{L^2(\oDe)}^2
\rp^{\frac{1}{2}}\ .
\ee
This allows, given a parameter
$\ka>0$, to define for any $\ph\in{\rm Fld}(X)$
the {\em large field regulator}
\be
G_\ka(X,\ph)\eqdef\exp\lp
\ka ||\ph||_{X,1,4}^2
\rp\ .
\ee
An important point is that $G_\ka(X,\cdot)$
is continuous on ${\rm Fld}(X)$.

\subsection{Some natural maps}
\label{mapsnatural}

Note that if $X_1\subset X_2$ are two polymers
then there is an obvious linear continuous restriction
map ${\rm Fld}(X_2)\rightarrow{\rm Fld}(X_1)$,
$\ph\mapsto \ph|_{X_1}$.
Indeed one first defines this projection from
$\widetilde{\rm Fld}(X_2)$ to $\widetilde{\rm Fld}(X_1)$.
Namely, it projects $\ph=(\ph_\De)_{\De\subset X_2}$
onto $(\ph_\De)_{\De\subset X_1}$. The gluing
conditions for the image are automatically satisfied if they
hold for the input $\ph$.

Now let $\ta$ be an isometry of Euclidean $\RR^3$ which leaves the lattice
$\ZZ^3$ globally invariant, and let $X$ be a polymer.
One has a natural Hilbert space isometry ${\rm Fld}(X)\rightarrow{\rm Fld}(\ta^{-1}(X))$,
$\ph\mapsto \ph\circ\ta$.
Indeed one first defines this map on elements $\ph=(\ph_\De)_{\De\subset X}
\in\widetilde{\rm Fld}(X)$ where each component is smooth on $\oDe$,
by ordinary composition with $\ta$. Then by density~\cite[Theorem 3.17]{Adams},
one extends it to a map $\widetilde{\rm Fld}(X)\rightarrow
\widetilde{\rm Fld}(\ta^{-1}(X))$.
Finally one takes the restriction to ${\rm Fld}(X)$ and corestriction
to ${\rm Fld}(\ta^{-1}(X))$, since the gluing conditions are preserved.

We will also need an additional map. Let $X\in{\rm Poly}_0$.
Then $LX$ is also in ${\rm Poly}_0$. Given $\ph\in{\rm Fld}(X)$
one can associate to it by a linear continuous map
an element $\ph_{L^{-1}}\in{\rm Fld}(LX)$ as follows.
First assume that $\ph=(\ph_\De)_{\De\subset X}\in\widetilde{\rm Fld}(X)$
is such that each $\ph_\De$ is smooth on $\oDe$. Then for each $\De\subset X$,
define $(\ph_\De)_{L^{-1}}(x)\eqdef L^{-[\ph]}\ph_\De(L^{-1}x)$
which is smooth in the interior of $L\De$.
Then for any unit box $\De'\subset L\De$ consider
the restriction
$(\ph_\De)_{L^{-1}}|_{\stackrel{\circ}{\De'}}$
to the interior of $\De'$. The collection of all such restrictions
for $\De'\subset L\De$ with $\De\subset X$ is by definition
the image of $\ph$ in $\widetilde{\rm Fld}(LX)$. Then extend the map, by
density, to all of $\widetilde{\rm Fld}(X)$.
Finally the wanted map is obtained by restriction to ${\rm Fld}(X)$ and corestriction
to ${\rm Fld}(LX)$, since the gluing conditions are easily seen
to be preserved.

\subsection{Gaussian measures}

Now given any polymer $X$, and using the standard theory
of Gaussian probability measures in Hilbert spaces~\cite{Bogachev,Skorohod},
it is not difficult to show that there exists a unique
Borel (with respect to
the $||.||_{\rm Fld(X)}$ norm topology) centered Gaussian
probability
measure $d\mu_{\Ga,X}$ on ${\rm Fld}(X)$ 
such that for any $x,y\in X$,
one has
\be
\int d\mu_{\Ga,X}(\ze)
\ \ze(x)\ze(y)=\Ga(x-y)
\label{covptwise}
\ee
where $\ze(x)$ and $\ze(y)$ are defined using the $C^2(X)$
realization of $\ze$.
In other words the covariance of $d\mu_{\Ga,X}$ is
the {\em fluctuation covariance} $\Ga$.

Indeed, one can define a continuous operator
$\tilde{S}:\widetilde{\rm Fld}(X)\rightarrow
\widetilde{\rm Fld}(X)$ as follows.
If $\ph=(\ph_\De)_{\De\subset X}\in\widetilde{\rm Fld}(X)$ has smooth components,
one defines its image
$\tilde{S}\ph=\lp(\tilde{S}\ph)_\De\rp_{\De\subset X}$
by letting for any $x\in\oDe$,
\be
(\tilde{S}\ph)_\De (x)\eqdef\sum_{\De'\subset X}\sum_{|\al|\le 4}
\int_{\De'} dy\ (-1)^{|\al|}
\partial^\al\Ga(x-y)\ \partial^\al\ph_{\De'}(y)\ .
\ee
It is easy to see that $\tilde{S}$ extends on all of $\widetilde{\rm Fld}(X)$
to a continuous
operator with norm bounded by
$\max_{|\al|\le 8} ||\partial^\al\Ga||_{L^\infty(\RR^3)}$. Clearly this operator
$\tilde{S}$ has its image contained in the closed subspace ${\rm Fld}(X)$.
It is also symmetric, and positive.
Now define the operator $S:{\rm Fld}(X)\rightarrow{\rm Fld}(X)$
by restriction and corestriction. It is easy to show that
\be
tr\ \tilde{S}=tr\ S= |X|\cdot\sum_{|\al|\le 4} (-1)^{|\al|}
\partial^{2\al}\Ga(0)\ .
\ee
As a result $S$ is a continuous symmetric positive trace class operator
on ${\rm Fld}(X)$, i. e., a covariance operator.
By the results in~\cite[Chapter 1]{Skorohod}, there exists
a unique centered Borel Gaussian probability measure $d\mu_{\Ga,X}$ on 
${\rm Fld}(X)$ such that for any $\ph_1,\ph_2\in{\rm Fld}(X)$,
\be
\int d\mu_{\Ga,X}(\ze)
\ (\ph_1,\ze)(\ph_2,\ze)=(\ph_1,S\ph_2)\ .
\label{covscalprod}
\ee
This equality also holds for $\ph_1,\ph_2$ more generally in
$\widetilde{\rm Fld}(X)$ and with
$S$ replaced by $\tilde{S}$.
It is not difficult to show that (\ref{covptwise}) follows from
(\ref{covscalprod}).
The uniqueness of Gaussian measures satisfying (\ref{covptwise})
is also easy. Indeed one has the uniqueness of Gaussian measures satisfying
(\ref{covscalprod}), see~\cite[Chapter 1]{Skorohod}.
Besides, consider the continuous
linear forms on $\widetilde{\rm Fld}(X)$ indexed by pairs $(\De,x)$
where $\De\subset X$ and $x\in\De$, obtained by evaluating at $x$
the $C^2(\De)$ image of the component $\ph_\De$ of a vector
$\ph\in\widetilde{\rm Fld}(X)$.
Let $\psi_{\De,x}\in\widetilde{\rm Fld}(X)$ 
be the corresponding vectors obtained by the Riesz representation theorem.
By the injectivity of the Sobolev embedding, it is clear that
the subspace generated by the vectors $\psi_{\De,x}$ is dense in
$\widetilde{\rm Fld}(X)$. The uniqueness then follows easily.

Finally,
note that if $X_1\subset X_2$ are two polymers,
then
the direct image measure of $d\mu_{\Ga,X_2}$, obtained by the restriction map
$\ph\mapsto\ph|_{X_1}$, coincides with $d\mu_{\Ga,X_1}$.

\subsection{Polymer activities}

Let $\KK$ denote either the (algebraic) field of real numbers
$\RR$ or that of complex numbers $\CC$. 
The main objects of study in this article are {\em polymer activities}
or {\em polymer amplitudes}. These are functions (or functionals)
$K(X,\cdot)$ from ${\rm Fld}(X)$ to $\KK$.
We will only consider functionals
which are $n_0$ times continuously differentiable in the sense of Frechet
between the {\em real} Banach spaces ${\rm Fld}(X)$ and
$\KK$~\cite[Chapter 2]{Berger},~\cite[Chapter VIII]{Dieudonne}.
Here $n_0$ is a nonnegative integer constant which we will actually
take to be $n_0=9$ as in~\cite{BMS}.

Now consider for any integer $n$, $0\le n\le n_0$,
the $\KK$-Banach space $\cLL_n({\rm Fld}(X),\KK)$
of $\RR$-multilinear continuous maps
$W:{\rm Fld}(X)^n\rightarrow\KK$ with the natural norm
\be
||W||_{\natural}
\eqdef\sup\limits_{\ph_1,\ldots,\ph_n\in {\rm Fld}(X)\backslash\{0\}}
\frac{|W(\ph_1,\ldots,\ph_n)|}
{||\ph_1||_{{\rm Fld}(X)}\ldots||\ph_n||_{{\rm Fld}(X)}}\ .
\ee
Inside it sits the space $\cLL_n({\rm Fld}(X),C^2(X),\KK)$ of $W$'s
for which the stronger norm
\be
||W||_{\sharp}
\eqdef\sup\limits_{\ph_1,\ldots,\ph_n\in {\rm Fld}(X)\backslash\{0\}}
\frac{|W(\ph_1,\ldots,\ph_n)|}
{||\ph_1||_{C^2(X)}\ldots||\ph_n||_{C^2(X)}}
\ee
is finite.
We indeed have for any $W\in \cLL_n({\rm Fld}(X),C^2(X),\KK)$
\be
||W||_{\natural}\le C_{\rm Sobolev}^n ||W||_{\sharp}\ .
\ee
It is easy to see that $\cLL_n({\rm Fld}(X),C^2(X),\KK)$
equipped with the sharp norm is a $\KK$-Banach space.
Let us denote by $C_\natural^{n_0}({\rm Fld}(X),\KK)$
the $\KK$-vector space of $\KK$-valued functionals $K(X,\cdot)$
defined on all of ${\rm Fld}(X)$, which are $n_0$ times
continuously Frechet differentiable in the usual
sense~\cite{Berger,Dieudonne}
with respect to the $||\cdot||_{{\rm Fld}(X)}$ topology.
We will also denote
the $n$-th Frechet differential at the point $\ph\in {\rm Fld}(X)$
of a polymer activity
$K(X,\cdot)$ by $D^n K(X,\ph)$.
Its evaluation at the sequence of vectors $f_1,\ldots,f_n$
of ${\rm Fld}(X)$
is
\be
D^n(X,\ph;f_1,\ldots,f_n)=
\left.
\frac{\partial^{n}}{\partial s_1\ldots\partial s_n}
K(X,\ph+s_1 f_1+\cdots+s_n f_n)
\right|_{s=0}\ ,
\ee
i.e., the corresponding directional or Gateau derivative.
We then define the space $C_\sharp^{n_0}({\rm Fld}(X),\KK)$
of all $K(X,\cdot)\in C_\natural^{n_0}({\rm Fld}(X),\KK)$
such that for all $\ph\in{\rm Fld}(X)$ and all integer $n$,
$0\le n\le n_0$, the differential
$D^n K(X,\ph)$ belongs to $\cLL_n({\rm Fld}(X),C^2(X),\KK)$,
and such that the maps $\ph\mapsto D^n K(X,\ph)$
are continuous from
$({\rm Fld}(X),||\cdot||_{{\rm Fld}(X)})$
to
$(\cLL_n({\rm Fld}(X),C^2(X),\KK),||\cdot||_\sharp)$.
From now on the only norm we will be considering for
differentials is the sharp one, therefore we will
omit the symbol from the norm notation.

Given a parameter $h>0$, a functional
$K(X,\cdot)\in C_\sharp^{n_0}({\rm Fld}(X),\KK)$
and a field $\ph\in{\rm Fld}(X)$, we define the {\em local norm}
\be
||K(X,\ph)||_h
\eqdef
\sum\limits_{0\le n\le n_0}
\frac{h^n}{n!}
||D^n K(X,\ph)||\ .
\ee
This allow us to define
the space $\cB_{h,G_\ka}^\KK(X)$ of all
$K(X)\in  C_\sharp^{n_0}({\rm Fld}(X),\KK)$
for which the norm
\be
||K(X)||_{h,G_\ka}
\eqdef\sup\limits_{\ph\in {\rm Fld}(X)}
G_\ka(X,\ph)^{-1}
||K(X,\ph)||_h
\ee
is finite.
Now one has the following easy proposition.

\begin{proposition}
For any $h,\ka>0$,
the normed $\KK$-vector space
\[
(\cB_{h,G_\ka}^{\KK}(X), ||\cdot||_{h,G_\ka})
\]
is complete.
\end{proposition}

Now we consider an arbitrary element $K=(K(X))_{X\in{\rm Poly}_0}$
in the product
\[
\prod\limits_{X\in{\rm Poly}_0}
\cB_{h,G_\ka}^{\KK}(X)\ ,
\]
and define the norm
\be
||K||_{h,G_\ka,\cA}
\eqdef\sup\limits_{\De\in{\rm Box}_0}
\sum\limits_{{X\in{\rm Poly}_0}\atop{X\supset\De}}
\cA(X)\ ||K(X)||_{h,G_\ka}
\ee
where $\cA$ is the previously defined large set regulator.
Given a parameter $h_\ast>0$ we also define
the {\em kernel semi-norm}
\be
\label{kernelsn}
|K|_{h_\ast,\cA}
\eqdef\sup\limits_{\De\in{\rm Box}_0}
\sum\limits_{{X\in{\rm Poly}_0}\atop{X\supset\De}}
\cA(X)\ ||K(X,0)||_{h_\ast}
\ee
where the differentials are taken at the point $\ph=0$
in each ${\rm Fld}(X)$.
We now introduce the notion of {\em calibrator}, it
is a new parameter $\gb>0$. We will use it
to set 
\be
h=c \gb^{-\frac{1}{4}}
\label{hgbdep}
\ee
for some fixed constant $c>0$
to be adjusted later. We will take
\be
h_\ast\eqdef L^{\frac{3+\ep}{4}}\ .
\ee

The space of all $K$ in the previous product space,
such that $||K||_{h,G_\ka,\cA}$
and $|K|_{h_\ast,\cA}$ are finite, is equipped
with the {\em calibrated norm}
\be
|||K|||_{\gb}\eqdef\max\lp
|K|_{h_\ast,\cA}, {\gb}^2
||K||_{h,G_\ka,\cA}
\label{calib_def}
\rp
\ee
and it is denoted by
$\cB\cB_{\gb}^{\KK}$.
So as to keep notations under control we only
emphasized the dependence on the calibrator $\gb$
which is the most important one in what follows.
One should keep in mind that the calibrated norm depends
on $\gb$ through the ${\gb}^2$ factor in front
of $||\cdot||_{h,G_\ka,\cA}$, but also through
the relation (\ref{hgbdep}) imposed between the $h$ parameter
and the calibrator $\gb$.
It is easy to see that $\cB\cB_{\gb}^{\KK}$ with
the norm $|||\cdot|||_{\gb}$, is a $\KK$-Banach space.

Now let $\ta$ be an isometry of Euclidean $\RR^3$ which
leaves the lattice $\ZZ^3$ globally invariant.
This transformation can be made to act on an element $K$ of
$\cB\cB_{\gb}^{\KK}$ by letting for any
$X\in{\rm Poly}_0$, and any
$\ph\in {\rm Fld}(X)$,
\be
(\ta K)(X,\ph)\eqdef K(\ta^{-1}(X),\ph\circ\ta)
\ee
where the
map
from ${\rm Fld}(X)$ to ${\rm Fld}(\ta^{-1}(X))$, given by
$\ph\mapsto \ph\circ\ta$ is the one defined in Section \ref{mapsnatural}.
We will only consider $\ta\in{\rm Transf}$ where
the set ${\rm Transf}$ is made of all translations by a vector
$m=(m_1,m_2,m_3)$ in $\ZZ^3$, together with the three orthogonal
reflections with respect to the coordinate planes respectively given
by the equations $x_1=0$, $x_2=0$ and $x_3=0$.
We also define a transformation
$K \mapsto K^-$ of $\cB\cB_{\gb}^{\KK}$
by letting $K^-(X,\ph)\eqdef K(X,-\ph)$.

The following lemma is an easy consequence of our previous
definitions for norms.

\begin{lemma}
The maps $K\mapsto \ta K$, for $\ta\in{\rm Transf}$,
as well as the map $K \mapsto K^-$, are Banach space
isometries of $\cB\cB_{\gb}^{\KK}$.
\end{lemma}

Thanks to this lemma we can finally define 
the main setting for a single RG map.
It is the space $\cB\cB\cS_{\gb}^{\KK}$
of all collections of polymer activities
$K\in\cB\cB_{\gb}^{\KK}$ such that
$K^-=K$ and for any $\ta\in{\rm Transf}$, $\ta K=K$.
By the previous lemma it is a closed subspace of
$\cB\cB_{\gb}^{\KK}$
and therefore a $\KK$-Banach space for the norm $|||\cdot|||_{\gb}$.
\begin{remark}
Note that all the calibrated norms, obtained for different values
of ${\gb}$, are equivalent. The underlying topological
vector spaces of the $\cB\cB_{\gb}^{\KK}$'s are therefore
the same.
\end{remark}
The RG map we are interested in is one from a domain
in $\RR\times\RR\times\cB\cB\cS_{\gb}^{\RR}$
for some values of the parameters
into another analogous triple-product space with a
slightly different value of ${\gb}$.
We will need complex versions of these spaces in order to obtain
Lipschitz contractive estimates with the least effort.
The global trajectory we construct in this article will
be obtained by a contraction mapping theorem
in a big Banach space of sequences $\cB\cB\cS\cS^\KK$ to be precisely
defined in Section \ref{sectBBSS} below.


\section{The Algebraic definition of the RG map}
\label{algebra}

In this section we provide all the formulae
which express the RG map studied in~\cite{BMS}.
We consider an input $(g,\mu,R)\in \CC\times\CC\times \cB\cB\cS^\CC$;
and we will give the algebraic definition for the output
$(g',\mu',R')$.
Recall that the latter have the form
\bea
g' & = &  L^\ep g-L^{2\ep} a(L,\ep) g^2+\xi_{g}(g,\mu,R)\ ,\\
\mu' & = &  L^{\frac{3+\ep}{2}} \mu+\xi_{\mu}(g,\mu,R)\ ,\\
R' &  = &  \cLL^{(g,\mu)}(R)+\xi_{R}(g,\mu,R)
\eea
where $a(L,\ep)$ has already been defined. We will therefore
provide the expressions for the $\xi$ remainders as well as
for $ \cLL^{(g,\mu)}(R)$.

\subsection{The local potentials}
\label{locpot}

For any $X\in{\rm Poly}_0$, any Borel set $Z\subset\RR^3$, and
any $\ph\in{\rm Fld}(X)$, we let
\be
V(X,Z,\ph)\eqdef
g\int_{Z\cap X} d^3 x\ :\ph(x)^4:_C 
+\mu \int_{Z\cap X} d^3 x\ :\ph(x)^2:_C\ . 
\label{Vpot}
\ee
We refer for instance to~\cite{GlimmJ,Salmhofer} for a discussion
of Wick ordering $:\bullet:_C$.
Otherwise the explicit expressions
\be
:\ph(x)^2:_C = \ph(x)^2-C(0)
\ee
and
\be
:\ph(x)^4:_C = \ph(x)^4-6C(0) \ph(x)^2 +3C(0)^2
\ee
may be used as definitions.
Note that in~\cite{BMS} the notation is simplified
to $V(Z,\ph)$ or even $V(Z)$ leaving the $\ph$ dependence
implicit. Here we prefer to keep everything explicit
including the first $X$ argument which allows one to keep track
of which space ${\rm Fld}(\cdot)$ the field $\ph$ lives in.
Also note that the function $\ph$ used in the integral formula
above is of course the $C^2(X)$ realization
of $\ph\in{\rm Fld}(X)$ via the embedding (\ref{C2Xembed}).
Another remark is that although we made the definition
sound quite general by allowing $Z$ to be any Borel
set, we will only need such $Z$'s which are complements
of the union of some $L^{-1}$-polymers in $X$.
Now define
\bea
g_L & \eqdef & L^{\ep} g\ ,\\
\mu_L & \eqdef & L^{\frac{3+\ep}{2}}\mu\ ,\\
C_{L^{-1}}(x) & \eqdef & L^{-2[\ph]} C(L^{-1}x)
\eea
and as in (\ref{Vpot})
let
\be
\tilde{V}(X,Z,\ph)\eqdef
g\int_{Z\cap X} d^3 x :\ph(x)^4:_{C_{L^{-1}}} 
+\mu \int_{Z\cap X} d^3 x :\ph(x)^2:_{C_{L^{-1}}} 
\ee
where Wick ordering is with respect to $C_{L^{-1}}$ instead of $C$.
Also let
\be
\tilde{V}_L(X,Z,\ph)\eqdef
g_L\int_{Z\cap X} d^3 x :\ph(x)^4:_C 
+\mu_L \int_{Z\cap X} d^3 x :\ph(x)^2:_C\ . 
\ee

\subsection{The $w$ kernels}
We now deal with the hidden variable $\mathbf{w}$.
Note that by construction the cutoff function $u_\ep$
satisfies $u_\ep(x)=0$ if $|x|_2\ge 1$ and a fortiori
if $|x|_\infty\ge 1$.
This implies that the fluctuation covariance $\Ga$ satisfies
$\Ga(x)=0$ if $|x|_\infty\ge L$.
Now we define $\mathbf{w}=\mathbf{w}_\ast=(w^{(1)},
w^{(2)},w^{(3)})$ to be a triple of real functions $w^{(p)}\in\cW_p$
where $\cW_p$, $p=1,2,3$, is the weighted $L^\infty$ space
$L^\infty(\RR^3, |x|_{\infty}^{\frac{3p}{2}} d^3 x)$.
Namely, $f\in \cW_p$ if and only if $f:\RR^3\rightarrow \RR$
is measurable and
\be
||f||_p\eqdef{\rm ess.}\sup\limits_{x\in\RR^3}
\lp
|x|_{\infty}^{\frac{3p}{2}} |f(x)|
\rp
\ee
is finite.
The $w$'s were constructed in~\cite[Lemma 5.9]{BMS}
by a Banach fixed point argument. We instead give them
explicitly, for $x\neq 0$, by
\bea
w^{(p)}(x) & \eqdef & \tilde{C}(x)^p-C(x)^p\\
 & = & \left[
C(x)+\int_0^1 \frac{dl}{l}\ l^{-\lp\frac{3-\ep}{2}\rp}
u_\ep\lp\frac{x}{l}\rp
\right]^p-C(x)^p\ .
\eea
From the last equation it is clear that $w^{(p)}(x)=0$
if $|x|_\infty\ge 1$.
Besides, since $u_\ep\ge 0$, for $\ep$ small
one has
\be
|w^{(p)}(x)|\le \tilde{C}(x)^p
=\frac{\vka_\ep^p}{|x|_2^{p\lp\frac{3-\ep}{2}\rp}}
\le \frac{\cO(1)}{|x|_\infty^{\frac{3p}{2}}} \ .
\ee
The fixed point property $\mathbf{w}=\mathbf{w}_\ast$
is embodied in the equation
\be
w^{(p)}(x)=v^{(p)}(x)+w_L^{(p)}(x)
\label{wfix}
\ee
for any $x\neq 0$, where we used the notation
\be
v^{(p)}(x)\eqdef C_L(x)^p-C(x)^p
\ee
and
\be
w_L^{(p)}(x)\eqdef L^{2p[\ph]} w^{(p)}(Lx)\ .
\ee
Equation (\ref{wfix}) trivially follows from the given definition.

\subsection{The renormalized expanded quadratic activity $Q$}
\label{theQ}

For $X\in{\rm Poly}_0$ and $\ph\in{\rm Fld}(X)$ we define
the activity $Q(X,\ph)$ as follows.
If $X$ is not ultrasmall we let
$Q(X,\ph)\eqdef 0$.
If $X$ is ultrasmall we introduce an associated
integration domain $\tilde{X}\subset \RR^3\times\RR^3$.
If $X$ is reduced to a single unit box $\De$,
we let $\tilde{X}\eqdef \De\times \De$.
If $X=\De_1\cup\De_2$ where the boxes $\De_1$ and $\De_2$ are
distinct but neighbouring, we let
\be
\tilde{X}\eqdef(\De_1\times\De_2)\cup(\De_2\times\De_1)\ .
\ee
We now write
\[
Q(X,\ph)\eqdef g^2 \int_{\tilde{X}} d^3 x\ d^3 y
\left\{
-24 w^{(3)}(x-y):(\ph(x)-\ph(y))^2:_C
\right.
\]
\[
-18 w^{(2)}(x-y):(\ph(x)^2-\ph(y)^2)^2:_C
\]
\be
\left.
+8 w^{(1)}(x-y):\ph(x)^3\ph(y)^3:_C
\right\}\ .
\ee
For reference, the Wick ordered expressions are explicitly given
by
\bea
:(\ph(x)-\ph(y))^2:_C & = & (\ph(x)-\ph(y))^2
-2C(0)+2C(x-y)\ ,\\
:(\ph(x)^2-\ph(y)^2)^2:_C & = & (\ph(x)^2-\ph(y)^2)^2
-4C(0)\ph(x)^2-4C(0)\ph(y)^2\nonumber\\
 & & +8C(x-y) \ph(x)\ph(y)+ 4C(0)^2-4C(x-y)^2\ ,\nonumber\\
 & & \ 
\eea
and
\bea
:\ph(x)^3\ph(y)^3:_C & = & \ph(x)^3\ph(y)^3
-3C(0) \ph(x)\ph(y)^3 -3C(0) \ph(x)^3\ph(y)\nonumber\\
 & & -9 C(x-y)\ph(x)^2\ph(y)^2+9 C(0)^2\ph(x)\ph(y)\nonumber\\
 & & +18 C(x-y)^2 \ph(x)\ph(y)+9 C(0) C(x-y) \ph(x)^2\nonumber\\
 & & +9 C(0) C(x-y) \ph(y)^2-9 C(0)^2 C(x-y)\nonumber\\
 & & -6 C(x-y)^3\ .
\eea

\subsection{Integration on fluctuations, reblocking and rescaling}
\label{section_expansion}

For any unit box $\De$ and fields $\ph,\ze\in{\rm Fld}(\De)$
we define
\be
P(\De,\ph,\ze)\eqdef e^{-V(\De,\De,\ph+\ze)}-e^{-\tilde{V}(\De,\De,\ph)}\ .
\label{defofP}
\ee
Now for any $X\in{\rm Poly}_0$ and $\ph\in{\rm Fld}(X)$
we let
\be
K(X,\ph)\eqdef Q(X,\ph) e^{-V(X,X,\ph)}+R(X,\ph)\ .
\ee
We also define
\be
R^\sharp(X,\ph)\eqdef \int d\mu_{\Ga,X}(\ze)
\ R(X,\ph+\ze)\ ,
\ee
as well as
\[
(\cS K)^\natural(X,\ph)\eqdef\int d\mu_{\Ga,LX}(\ze)
\left\{
\sum_{{M,N}\atop{M+N\ge 1}}\frac{1}{M!N!}
\sum_{(\De_1,\ldots,\De_M),(X_1,\ldots,X_N)}
\right.
\]
\[
\exp\left[
-\tilde{V}\lp
LX, LX\backslash\lp
\lp\cup_{i=1}^{M} \De_i\rp
\cup
\lp\cup_{j=1}^{N} X_j\rp
\rp, \ph_{L^{-1}}\rp
\right]
\]
\be
\left.
\times
\prod\limits_{i=1}^{M}
P\lp
\De_i,\ph_{L^{-1}}|_{\De_i}, \ze|_{\De_i}
\rp
\times
\prod\limits_{j=1}^{N}
K\lp
X_j,\ph_{L^{-1}}|_{X_j}+ \ze|_{X_j}
\rp
\begin{array}{c}
 \\
 \\
 \\
 \\
 
\end{array}
\right\}
\label{defofSKnat}
\ee
where the sum over sequences $(\De_1,\ldots,\De_M)$ and
$(X_1,\ldots,X_N)$ is subjected to the following
conditions:
\begin{enumerate}
\item
The $\De_i$ are {\em distinct} boxes in ${\rm Box}_0$.
\item
The $X_j$ are {\em disjoint} polymers in ${\rm Poly}_0$.
\item
None of the $\De_i$ is contained in an $X_j$.
\item
The $L$-closure of the union of all the $\De_i$ and the $X_j$
is exactly the set $LX$.
\end{enumerate}

\begin{remark}
Note that since the $X_j$ are {\em closed} polymers,
the disjointness condition means that they cannot touch
each other and have to be at least $1$ apart in
$|\cdot|_\infty$ distance.
However, the $\De_i$ are allowed to touch each other
or an $X_j$, by sharing no more than a boundary component.
Also note that by hypothesis, $X$ and therefore $LX$ is connected.
This rules out situations where for instance the $(X_j)$ sequence
would be empty, and the $(\De_i)$ sequence would be made of two
boxes very far apart.
\end{remark}

\subsection{Preparations for the extraction}
\label{prepaext}

As a preparation for the crucial so called extraction step
we need to introduce for any $X\in{\rm Poly}_0$
the quantities denoted by $\tilde{\al}_0(X)$,
$\tilde{\al}_2(X)$, $\tilde{\al}_{2,\mu}(X)$
for $\mu=1,2,3$, and $\tilde{\al}_4(X)$.
These are by definition all set to zero if $X$ is large.
Now if $X$ is small one lets
\be
\tilde{\al}_0(X)\eqdef\frac{e^{\tilde{V}(X,X,0)}}{|X|}
R^\sharp(X,0)\ ,
\ee
\be
\tilde{\al}_2(X)\eqdef\frac{e^{\tilde{V}(X,X,0)}}{2|X|}
\left[
D^2(R^\sharp)(X,0;1,1)
+R^\sharp(X,0)
D^2\tilde{V}(X,X,0;1,1)
\right]
\ee
where the last two arguments of the differentials
are given by the constant function equal to 1, seen
as an element of ${\rm Fld}(X)$.
We also let for $\mu=1,2,3$,
\bea
\tilde{\al}_{2,\mu}(X) & \eqdef & \frac{e^{\tilde{V}(X,X,0)}}{|X|}
\bigg[
D^2(R^\sharp)(X,0;1,\De_X x_\mu)\nonumber\\
 & & +R^\sharp(X,0) D^2\tilde{V}(X,X,0;1,\De_X x_\mu)
\bigg]
\eea
where $\De_X x_\mu$ means the function
\[
x\mapsto
x_\mu-\frac{1}{|X|}
\lp \int_X d^3 y \ y_\mu
\rp
\]
the deviation from average of the coordinate function $x_\mu$
on the polymer $X$,
again seen
as an element of ${\rm Fld}(X)$.
Finally one lets
\bea
\tilde{\al}_4(X) & \eqdef & \frac{e^{\tilde{V}(X,X,0)}}{24|X|}
\Bigg[D^4(R^\sharp)(X,0;1,1,1,1)\nonumber\\
 & & +6 D^2(R^\sharp)(X,0;1,1)D^2\tilde{V}(X,X,0;1,1)
\nonumber\\
 & & +R^\sharp(X,0) D^4\tilde{V}(X,X,0;1,1,1,1)\nonumber\\
 & & +3 R^\sharp(X,0)\lp D^2\tilde{V}(X,X,0;1,1)
\rp^2\Bigg]\ .
\eea
Now given $Z\in{\rm Poly}_0$, and $x\in\RR^3$ we define
\bea
\al_0(Z,x) & \eqdef & \sum_{X\ {\rm small},\ \bar{X}^L=LZ}
\tilde{\al}_0 (X) L^3 \bbbone_{L^{-1}X}(x)\ ,\\
\al_2(Z,x) & \eqdef & \sum_{X\ {\rm small},\ \bar{X}^L=LZ}
\tilde{\al}_2 (X) L^{\frac{3+\ep}{2}} \bbbone_{L^{-1}X}(x)\ ,\\
\al_{2,\mu}(Z,x) & \eqdef & \sum_{X\ {\rm small},\ \bar{X}^L=LZ}
\tilde{\al}_{2,\mu} (X) L^{\frac{1+\ep}{2}} \bbbone_{L^{-1}X}(x)\ ,\\
\al_4(Z,x) & \eqdef & \sum_{X\ {\rm small},\ \bar{X}^L=LZ}
\tilde{\al}_4 (X) L^\ep \bbbone_{L^{-1}X}(x)
\eea
where again $\mu=1,2,3$, and $\bbbone_{L^{-1}X}$ denotes the sharp
characteristic function of the set $L^{-1}X$.
Note that these quantities vanish if $Z$ is not small
or if $x\notin Z$.

Now choose some reference box $\De_0\in{\rm Box}_0$.
We define
\bea
\al_0 & \eqdef & L^3
\sum_{X\ {\rm small},\ X\supset\De_0} \tilde{\al}_0 (X)\ ,\\
\al_2 & \eqdef & L^{\frac{3+\ep}{2}}
\sum_{X\ {\rm small},\ X\supset\De_0} \tilde{\al}_2 (X)\ ,\\
\al_4 & \eqdef & L^\ep
\sum_{X\ {\rm small},\ X\supset\De_0} \tilde{\al}_4 (X)\ .\\
\eea

Note that the latter do not depend on the choice of $\De_0$
because of the translational invariance imposed on polymer
activities in Section \ref{functspace}. Also note that
in~\cite[Equation 4.44]{BMS}
the quantities
\be
\al_{2,\mu}  \eqdef  L^{\frac{1+\ep}{2}}
\sum_{X\ {\rm small},\ X\supset\De_0} \tilde{\al}_{2,\mu} (X)
\ee
for $\mu=1,2,3$, were also defined. However, again by the conditions
imposed on polymer
activities in Section \ref{functspace}, it is easy to
see that the latter always vanish.
In other words, the RG flow does not create $\ph\partial\ph$
terms in the effective potential.

After one has defined
\be
b(L,\ep)\eqdef 48 \int_{\RR^3} d^3 x
\ v^{(3)}(x)\ ;
\ee
one can at last give some of the outputs of the RG map.
Namely, one poses
\bea
\xi_g(g,\mu,R) & \eqdef & -\al_4\ ,\\
\xi_\mu(g,\mu,R) & \eqdef & -\lp
L^{2\ep} b(L,\ep) g^2+\al_2+6 C(0)\al_4
\rp
\eea
as definition of the first two remainder terms.
At this point, the new couplings are defined via
\bea
g' & \eqdef &  L^\ep g-L^{2\ep} a(L,\ep) g^2+\xi_{g}(g,\mu,R)\ ,\\
\mu' & \eqdef &  L^{\frac{3+\ep}{2}} \mu+\xi_{\mu}(g,\mu,R)\ .
\eea
What remains is $\cLL^{(g,\mu)}(R)$, $\xi_{R}(g,\mu,R)$
and their combination $R'$.

\subsection{The linear map for $R$}

In order to define the linear part $\cLL^{(g,\mu)}(R)$
which was denoted by $R_{\rm linear}$ in~\cite{BMS},
we need to introduce two polymer activities.
For $X\in{\rm Poly}_0$, and $\ph\in{\rm Fld}(X)$,
we let $\tilde{F}_R(X,\ph)\eqdef 0$ if $X$ is large;
otherwise we let
\[
\tilde{F}_R(X,\ph)\eqdef \int_X d^3 x\Bigg[
\tilde{\al}_4(X) \ph(x)^4+
\tilde{\al}_2(X) \ph(x)^2
\]
\be
+\sum_{\mu=1}^{3} \tilde{\al}_{2,\mu}(X) \ph(x) \partial_\mu\ph(x)
+\tilde{\al}_0(X)
\Bigg]\ .
\ee
Regardless of whether $X$ is small or not, we also
let
\be
J(X,\ph)\eqdef R^\sharp(X,\ph)-\tilde{F}_R(X,\ph)
e^{-\tilde{V}(X,X,\ph)}\ .
\ee

The previous complicated definitions of the
$\tilde{\al}_{\ldots}(X)$
had no other purpose but to secure the following
{\em normalization conditions}. For any small polymer $X$,
and for $\mu=1,2,3$,
one needs
\bea
J(X,\ph) & = & 0\ ,\\
D^2 J(X,0;1,1) & = & 0\ ,\\
D^2 J(X,0;1,\De_X x_\mu) & = & 0\ ,\\
D^4 J(X,0;1,1,1,1) & = & 0\ .
\eea
Note that one would have equivalent conditions if one replaced
the function $\De_X x_\mu$ simply by the coordinate function $x_\mu$.
These normalization conditions are the analog
in the present setting of the BPHZ substraction prescription
(see e.g.~\cite{Rivass}).
They are the main reason why the map
$\cLL^{(g,\mu)}(\cdot)$ we are about to
define is contractive.

Now given $X\in{\rm Poly}_0$, and $\ph\in{\rm Fld}(X)$,
and using constrained sums over polymers $Y\in{\rm Poly}_0$,
we define
\[
\cLL^{(g,\mu)}(R)(X,\ph)\eqdef
\sum_{Y\ {\rm small},\ \bar{Y}^L=LX}
J(Y,\ph_{L^{-1}}|_Y) e^{-\tilde{V}_L(X,X\backslash L^{-1}Y,\ph)}
\]
\be
+\sum_{Y\ {\rm large},\ \bar{Y}^L=LX}
R^\sharp(Y,\ph_{L^{-1}}|_Y)
e^{-\tilde{V}_L(X,X\backslash L^{-1}Y,\ph)}\ .
\label{deflinmap}
\ee

\subsection{The extraction proper}

Given  $X\in{\rm Poly}_0$, and $x\in\RR^3$,
we define the function $f_Q^{(4)}(X,x)$ as follows.

\noindent{\bf First case: }
If $X$ is given by a single box $\De\in{\rm Box}_0$, and if
$x$ lies in the interior of $\De$, we let
\be
f_Q^{(4)}(X,x)\eqdef\int_{\De} d^3 y
\ v^{(2)}(x-y)\ .
\ee

\noindent{\bf Second case: }
If $X$ is given by the union of two distinct neighbouring boxes
$\De_1,\De_2\in{\rm Box}_0$, and if
$x$ lies in the interior of say $\De_1$, we let
\be
f_Q^{(4)}(X,x)\eqdef\int_{\De_2} d^3 y
\ v^{(2)}(x-y)\ .
\ee

\noindent{\bf Third case: }
If none of the first two cases apply, we simply
let $f_Q^{(4)}(X,x)\eqdef 0$.

One can in the same manner define a function $f_Q^{(2)}(X,x)$
using $v^{(3)}$ instead of $v^{(2)}$, as well as
a function $f_Q^{(0)}(X,x)$
using $v^{(4)}$ which is given by $v^{(4)}(z)\eqdef C_L(z)^4-C(z)^4$.

Now let $X\in{\rm Poly}_0$, and $Z$ be a Borel set in $\RR^3$,
and define
\be
F_{0,Q}(X,Z)\eqdef 12 g_L^2 \int_{Z} d^3 x
f_Q^{(0)}(X,x)
\ee
as well as
\be
F_{0,R}(X,Z)\eqdef \int_{Z} d^3 x
\left\{
\al_0(X,x)+C(0) \al_2(X,x)+3C(0)^2 \al_4(X,x)
\right\}
\ee
and
\be
F_{0}(X,Z)\eqdef F_{0,Q}(X,Z)+F_{0,R}(X,Z)\ .
\ee

If in addition one has a polymer $Y\in{\rm Poly}_0$, and
a field $\ph\in{\rm Fld}(Y)$, one can also define
\[
F_{1,Q}(X,Y,Z,\ph)\eqdef
36 g_L^2 \int_{Z\cap Y} d^3 x
:\ph(x)^4:_C f_Q^{(4)}(X,x)
\]
\be
+48 g_L^2 \int_{Z\cap Y} d^3 x
:\ph(x)^2:_C f_Q^{(2)}(X,x)
\ee
as well as
\[
F_{1,R}(X,Y,Z,\ph)\eqdef\int_{Z\cap Y} d^3 x
\left\{\al_4(X,x) :\ph(x)^4:_C\right.
\]
\[
+\sum\limits_{\mu=1}^{3}  \al_{2,\mu}(X,x)
:\ph(x)\partial_\mu\ph(x):_C
\]
\be
\left.
+\lp
\al_2(X,x)+6 C(0) \al_4(X,x)
\rp :\ph(x)^2:_C
\right\}
\ee
where $:\ph(x)\partial_\mu\ph(x):_C$ reduces to $\ph(x)\partial_\mu\ph(x)$.
We finally need
\be
F_{1}(X,Y,Z,\ph)\eqdef F_{1,Q}(X,Y,Z,\ph)+F_{1,R}(X,Y,Z,\ph)\ ,
\ee
and
\be
F(X,Y,Z,\ph)\eqdef F_{0}(X,Z)+F_{1}(X,Y,Z,\ph)\ .
\ee
As before the $Y$ argument is for keeping track of
which ${\rm Fld}(\cdot)$ does $\ph$ live in.
The $Z$ defines the domain of integration.
The new argument $X$, is here to indicate
that the $F$'s are {\em local counterterms}
for a polymer activity which originally lived on $X$.

Now given $X\in{\rm Poly}_0$, and $\ph\in{\rm Fld}(X)$, we let
\bea
\tilde{K}(X,\ph) & \eqdef & (\cS K)^\natural(X,\ph)-e^{-\tilde{V}_L(X,X,\ph)}
\times
\sum_{N\ge 1}\frac{1}{N!}\sum\limits_{(Y_1,\ldots,Y_N)}
\nonumber\\
 & & \prod\limits_{i=1}^{N}
\left[
\exp\lp
F\lp Y_i,Y_i,Y_i,\ph|_{Y_i}
\rp
\rp-1
\right]
\eea
where the sum is over all sequences of {\em distinct} polymers
$Y_i\in{\rm Poly}_0$ whose union is equal to $X$.

Again given $X\in{\rm Poly}_0$, a Borel set $Z$, and a field
$\ph\in{\rm Fld}(X)$, we define
\be
V_F(X,Z,\ph)\eqdef\sum_{{\De\in{\rm Box}_0}\atop
{\stackrel{\circ}{\De}\subset Z\cap X}}
\left[
\tilde{V}_L\lp\De,\De,\ph|_\De\rp
-\sum_{{Y\in{\rm Poly}_0}\atop{Y\supset\De}}
F\lp Y,\De,\De,\ph|_\De\rp
\right]\ .
\label{renormpot}
\ee
Mind the inclusion condition only on the interior
$\stackrel{\circ}{\De}$ of $\De$.

Then for $X\in{\rm Poly}_0$, and $\ph\in{\rm Fld}(X)$,
we let
\bea
\tilde{\cE}(X,\ph) & \eqdef & \sum_{M\ge 1,\ N\ge 0}\frac{1}{M!N!}
\sum\limits_{(X_1,\ldots,X_M),\ (Z_1,\ldots,Z_N)}\nonumber\\
 & & \exp\lp-V_F\lp
X,X\backslash\lp\cup_{i=1}^{M} X_i\rp,\ph
\rp\rp
\times\prod\limits_{i=1}^{M} \tilde{K}\lp X_i,\ph|_{X_i}\rp
\nonumber\\
 & & \times\prod\limits_{j=1}^{N}
\left[
\exp\lp
-F\lp Z_j,Z_j,Z_j\backslash\lp\cup_{i=1}^{M} X_i\rp,
\ph|_{Z_j}\rp
\rp-1
\right]\nonumber\\
 & & \ 
\eea
with the following
conditions imposed on the $X_i$ and $Z_j$:
\begin{enumerate}
\item
The $X_i$ and $Z_j$ are polymers in ${\rm Poly}_0$.
\item
The $X_i$ are {\em disjoint}.
\item
The $Z_j$ are {\em distinct}.
\item
Every $Z_j$ has a nonempty intersection, be it by an edge or a corner,
with $\cup_{i=1}^{M} X_i$.
\item
Every $Z_j$ has a nonempty intersection with
$X\backslash\lp\cup_{i=1}^{M} X_i\rp$.
\item
The union of all the $X_i$ and $Z_j$ is exactly the given
polymer $X$.
\end{enumerate}

\begin{remark}
We emphasized the condition on the interior
of $\De$ in (\ref{renormpot}),
and the weak notion of intersection in items (4) and (5) above,
since these are the notable modifications to make
on the treatment of~\cite[Section 4.2]{BDH} in order to account
for the {\em closed} polymers used in~\cite{BMS} and here.
The overlap connectedness in~\cite[Section 4.2]{BDH}
is automatically implied
by item (6) above and the connectedness of the set $X$ which
is assumed a priori. Also note that this notion was
defined in~\cite[Section 4.2]{BDH} based on the idea of having
a full box in common, whereas here a nonempty intersection
by a boundary component already counts as an overlap.
Finally note that if $M\ge 2$ then one needs to have $N\ge 1$;
this is because the $X_i$ are forced
to be at least $1$ apart with respect to the
$|\cdot|_\infty$ distance, and they need a bridge
of $Z_j$'s joining them.
\end{remark}

Now given $X\in{\rm Poly}_0$, and $\ph\in{\rm Fld}(X)$, we let
\be
\cE(X,\ph)\eqdef\tilde{\cE}(X,\ph)
\times
\exp\left[
-\sum_{{\De\in{\rm Box}_0}\atop{\De\subset X}}
\sum_{{Y\in{\rm Poly}_0}\atop{Y\supset\De}} F_0(Y,\De)
\right]\ .
\ee

Finally we define $Q'(X,\ph)$ in exactly the same way
as $Q(X,\ph)$ in Section \ref{theQ}
but using the new coupling $g'$ obtained in Section \ref{prepaext}
instead of the old one $g$.
Likewise we need a potential $V'(X,Z,\ph)$
defined in the exact same manner as $V(X,Z,\ph)$
in Section \ref{locpot} using the new couplings
$g',\mu'$ instead of $g,\mu$.
At last one can give the output $R'$ of the RG map
defined for any $X\in{\rm Poly}_0$, and $\ph\in{\rm Fld}(X)$ by
\be
R'(X,\ph)\eqdef\cE(X,\ph)-Q'(X,\ph) e^{-V'(X,X,\ph)}\ .
\ee
In somewhat of a roundabout manner, the definition of
the $\xi_R$ remainder
is then
\be
\xi_R(g,\mu,R)(X,\ph)\eqdef R'(X,\ph)
-\cLL^{(g,\mu)}(R)(X,\ph)\ .
\ee

The algebraic definition of the RG map is now complete.
Note that the Frechet differentiability of the output
polymer activities, the justification of the measurability
of the integrations over $\ze$, follow once the proper estimates
are established because of the algebraic nature of the
operations used in this section. These estimates have been
provided in ~\cite[Section 5]{BMS}, and their result
is summarized in Theorem \ref{BMSestimates} below.


\section{The dynamical system construction}
\label{sectBBSS}

The RG map for which the defining formulae were
given in the previous section is $(g,\mu,R)\mapsto(g',\mu',R')$ where
\be
\left\{
\begin{array}{lll}
g' & = & L^\ep g-L^{2\ep} a(L,\ep) g^2+\xi_{g}(g,\mu,R)\ ,\\
\mu' & = & L^{\frac{3+\ep}{2}} \mu+\xi_{\mu}(g,\mu,R)\ ,\\
R' & = & \cLL^{(g,\mu)}(R)+\xi_{R}(g,\mu,R)\ .
\end{array}
\right.
\label{recursion}
\ee
Our aim is to construct a double-sided sequence
$s=(g_n,\mu_n,R_n)_{n\in\ZZ}$ which solves this recursion
and such that
$\lim\limits_{n\rightarrow -\infty}
(g_n,\mu_n,R_n)=(0,0,0)$
the Gaussian ultraviolet fixed point, and
$\lim\limits_{n\rightarrow +\infty}
(g_n,\mu_n,R_n)=(g_\ast,\mu_\ast,R_\ast)$
the BMS nontrivial infrared fixed point~\cite{BMS}.
We proceed as follows.
We will simply write $a$ for the coefficient $a(L,\ep)>0$.
We also take $\ep>0$ small enough so that
$L^\ep\in ]1,2[$. Recall that $\gbs=\frac{L^\ep -1}{L^{2\ep}a}>0$
and consider the
function
\be
\begin{array}{rl}
f: & [0,\gbs]\rightarrow [0,\gbs] \\
 & x\mapsto f(x)=L^\ep x-L^{2\ep}a x^2\ .
\end{array}
\ee
It is trivial to see that $f$ is a strictly increasing diffeomorphism
of $[0,\gbs]$; it is also strictly concave. The only fixed points
are $0$ and $\gbs$, and $f(x)>x$ in the interval $]0,\gbs[$.
Given $\omz\in]0,1[$, there is a unique double-sided sequence
$(\gb_n)_{n\in\ZZ}$ in $]0,\gbs[^\ZZ$
such that $\gb_0=\omz\gbs$, and for any
$n\in\ZZ$, $\gb_{n+1}=f(\gb_n)$. This sequence is strictly increasing
from $0$ when $n\rightarrow -\infty$, to $\gbs$ when $n\rightarrow +\infty$.
We call $\gb_0$ the coupling at unit scale.
Once it is chosen it defines the sequence $({\gb}_ n)_{n\in\ZZ}$
completely. Moreover, if one ignores the remainder terms $\xi$
in (\ref{recursion}) then the renormalization group recursion is solved
by the approximate sequence
$\bar{s}\eqdef({\gb}_n,0,0)_{n\in\ZZ}$.
The true trajectory will be constructed
in such a way that $g_0={\gb}_0$,
and 
via the construction of the deviation sequence
$\de s=(\de g_n,\mu_n,R_n)_{n\in\ZZ}$
with respect to the approximate sequence $\bar{s}$.
Using the notation $\de g_n=g_n-{\gb}_n$, the new recursion which
is equivalent to (\ref{recursion}) that we have to solve is
\be
\left\{
\begin{array}{lll}
\de g_{n+1} & = & f'(\gb_n) \de g_n
+\left[-L^{2\ep} a\ \de g_n^2\ +
\xi_{g}(\gb_n+\de g_n,\mu_n,R_n)\right]\ ,\\
\mu_{n+1} & = & L^{\frac{3+\ep}{2}} \mu_n+
\xi_{\mu}(\gb_n+\de g_n,\mu_n,R_n)\ ,\\
R_{n+1} & = & \cLL^{(\gb_n+\de g_n,\mu_n)} (R_n)+
\xi_{R}(\gb_n+\de g_n,\mu_n,R_n)\ .
\end{array}
\right.
\label{deltarecur}
\ee
The boundary conditions we will need can roughly be stated as:
\begin{itemize}
\item
$\de g_0=0$.
\item
$\mu_n$ does not blow up when $n\rightarrow+\infty$.
\item
$R_n$ does not blow up when $n\rightarrow-\infty$.
\end{itemize}
Also note the behavior of the linear parts of (\ref{deltarecur}) :
\begin{itemize}
\item
When $n\rightarrow +\infty$, $f'({\gb}_n)\rightarrow 2-L^\ep<1$,
i.e., one has a deamplification.
\item
When $n\rightarrow -\infty$, $f'({\gb}_n)\rightarrow L^\ep>1$,
i.e., one has an amplification.
\item
One always has an amplification in the `relevant'
$\mu$ or mass direction.
\item
Once the RG map has been properly defined, one can arrange
to always have a deamplification in the `irrelevant' $R$ direction.
\end{itemize}

Based on these observations, it is natural using the standard
method
of associated `discrete integral equations',
used for instance in~\cite{Irwin},
to rewrite the system
(\ref{deltarecur}) as
\bea
\lefteqn{\forall n>0,} \nonumber\\
 & &
\de g_{n}= f'(\gb_{n-1}) \de g_{n-1}
+\left[-L^{2\ep} a\ \de g_{n-1}^2\ +
\xi_{g}(\gb_{n-1}+\de g_{n-1},\mu_{n-1},R_{n-1})\right]\ ,
\nonumber\\
 & & {\ }\\
\lefteqn{\forall n<0,} \nonumber\\
 & &
\de g_{n}= \frac{1}{f'(\gb_n)} \de g_{n+1}
-\frac{1}{f'(\gb_n)}\left[-L^{2\ep} a\ \de g_{n}^2\ +
\xi_{g}(\gb_{n}+\de g_{n},\mu_{n},R_{n})\right]\ ,\\ 
\lefteqn{\forall n\in\ZZ,} \nonumber\\
 & &
\mu_n=L^{-\lp\frac{3+\ep}{2}\rp} \mu_{n+1}
-L^{-\lp\frac{3+\ep}{2}\rp} 
\xi_{\mu}(\gb_{n}+\de g_{n},\mu_{n},R_{n})\ ,\\
\lefteqn{\forall n\in\ZZ,} \nonumber\\
 & &
R_{n}=\cLL^{(\gb_{n-1}+\de g_{n-1},\mu_{n-1})} (R_{n-1})+
\xi_{R}(\gb_{n-1}+\de g_{n-1},\mu_{n-1},R_{n-1})\ ,
\eea
and iterate, i.e., replace the linear term occurences
of the dynamical variables $\de g$, $\mu$, $R$, in terms
of the analogous equations for $n-1$ or $n+1$, and repeat
ad nauseam until one hits a boundary condition.
In sum, the true sequence we are seeking will be
constructed as a fixed point of a map $s\mapsto s'$
or rather $\de s\mapsto \mF(\de s)$ which
to a sequence $\de s=(\de g_n,\mu_n,R_n)_{n\in\ZZ}$
associates the new sequence
$\mF(\de s)=(\de g'_n,\mu'_n,R'_n)_{n\in\ZZ}$
which is given as follows.
\begin{definition}
\label{defbigmap}
{\bf The map on sequences}

Leaving the issue of convergence for later,
the defining formulae for the map $\mF$ are :
\bea
\lefteqn{\de g'_0\eqdef 0\ ,}\\
 & & \nonumber\\
\lefteqn{\forall n>0,} \nonumber\\
 & & \de g'_n\eqdef\sum\limits_{0\le p<n}
\lp
\prod\limits_{p<j<n} f'(\gb_j)
\rp
\left[-L^{2\ep} a\ \de g_p^2\ +
\xi_{g}(\gb_p+\de g_p,\mu_p,R_p)\right]\ ,\nonumber\\
 & & \label{backforg}\\
\lefteqn{\forall n<0,} \nonumber\\
 & & \de g'_n\eqdef -\sum\limits_{n\le p<0}
\lp
\prod\limits_{n\le j\le p} \frac{1}{f'(\gb_j)}
\rp
\left[-L^{2\ep} a\ \de g_p^2\ +
\xi_{g}(\gb_p+\de g_p,\mu_p,R_p)\right]\ ,\nonumber\\
 & & \label{forwforg}\\
\lefteqn{\forall n\in\ZZ,} \nonumber\\
 & & \mu'_n\eqdef -\sum\limits_{p\ge n}
L^{-\lp\frac{3+\ep}{2}\rp(p-n+1)}
\ \xi_{\mu}(\gb_p+\de g_p,\mu_p,R_p)\ ,\label{forwformu}
\eea
and finally
\bea
\lefteqn{\forall n\in\ZZ,} \nonumber\\
 & & R'_n\eqdef\sum\limits_{p<n}
\cLL^{(\gb_{n-1}+\de g_{n-1},\mu_{n-1})}
\circ
\cLL^{(\gb_{n-2}+\de g_{n-2},\mu_{n-2})}
\circ\cdots\qquad\qquad \nonumber\\
 & &
\cdots\circ\cLL^{(\gb_{p+1}+\de g_{p+1},\mu_{p+1})}
\lp\xi_{R}(\gb_p+\de g_p,\mu_p,R_p)\rp\ .
\label{backforR}
\eea
where the composition $\circ$ is of course with respect to the
$R$ argument.
\end{definition}

We now come to the definition of the space in which the
deviation sequences $\de s$ will live.
Let us introduce as in~\cite{BMS} the exponent drops
$\de\in [0,\frac{1}{6}]$ and $\et\in[0,\frac{3}{16}]$
which will be fixed later.
We will also define for $n\in \ZZ$
\be
e_n\eqdef\left\{
\begin{array}{ll}
1 & {\rm if}\ n\le 0\ ,\\
\frac{3}{2} & {\rm if}\ n \ge 1\ .
\end{array}
\right.
\label{gexponent}
\ee
Now we define the big Banach space of sequences
\be
\cB\cB\cS\cS^\KK
\subset
\prod\limits_{n\in\ZZ}
\lp \KK\times\KK\times
\cB\cB\cS_{{\gb}_n}^\KK
\rp
\ee
whose elements are all deviation
sequences $\de s=(\de g_n,\mu_n,R_n)_{n\in\ZZ}$
for which the quadruple norm
\be
||||\de s||||\eqdef
\sup\limits_{n\in\ZZ}\lp
\max\left\{
|\de g_n| {\gb}_n^{-e_n},
|\mu_n| {\gb}_n^{-(2-\de)},
|||R_n|||_{{\gb}_n} {\gb}_n^{-(\frac{11}{4}-\et)}
\right\}
\rp
\ee
is bounded and such that $\de g_0=0$.
Note that the approximate sequence $\bar{s}$ itself does not
belong to $\cB\cB\cS\cS^\KK$
which somewhat plays the role of a tangent space
around it.
As an easy consequence of our definitions one has
the following proposition.
\begin{proposition}
The space
\[
\lp
\cB\cB\cS\cS^\KK,
||||\cdot||||
\rp
\]
is complete.
\end{proposition}


\section{The BMS estimates on a single RG step}
\label{sectest}

The estimates in~\cite[Section 5]{BMS}, slightly modified
for the needs of the present construction, can be summarized
by Theorem \ref{BMSthm} below.
Before stating the theorem one can give a brief description
of the main ideas
behind the estimates of~\cite[Section 5]{BMS}.
Given some a priori hypotheses on the size of the input $g,\mu,R$
of the RG map, the goal is to prove estimates on the output $g',\mu',R'$.
The size of these variables is typically measured in powers of the $\phi^4$
coupling $g$. However the latter is a dynamical variable of the problem,
and in order to avoid a vicious circle one uses instead powers of
a predetermined approximation $\bar{g}$ which we have called {\em the
calibrator}.  The true value of $g$ is allowed to float in a small
complex ball centred on $\bar{g}$. 
In~\cite[Equation 5.1]{BMS} this calibrator 
is taken equal to the approximate fixed point value which we denoted here by
$\gbs$ and which is of order $\ep$.
Grosso modo the main purpose of~\cite[Section 5]{BMS} is to show that
provided $\mu$ is of order
$\bar{g}^2$, and $R$ is of order $\bar{g}^3$,
then the linear map $\cLL^{(g,\mu)}$
is contractive in the $R$ direction, and the remainder $\xi_R$
remains of order $\bar{g}^3$. 
In fact, for technical reasons, the exponents are slightly altered and
a more precise statement would be: provided $\mu$ is of order
$\bar{g}^{2-\de}$, and $R$ is of order $\bar{g}^{\frac{11}{4}-\et}$,
then the linear map $\cLL^{(g,\mu)}$
is contractive in the $R$ direction, and the remainder $\xi_R$
remains of order $\bar{g}^{\frac{11}{4}}$. Here $\de$ and $\et$ are small
nonnegative discrepancies. 
A nice feature of
the estimates~\cite[Equation 5.1]{BMS} is that they allow a bound
on the output $\xi_R$ which is strictly better than the one on the input
$R$, when $\et>0$. This is required in the subsequent dynamical system
construction,
for an effective use of the splitting
$R'=\cLL^{(g,\mu)}(R)+\xi_R$.

Two norms are required to measure the $R$
coordinate. The first is the \emph{kernel} semi-norm
$|\cdot|_{h_{*},\mathcal{A}}$ defined in (\ref{kernelsn}). 
This norm detects the
true power $g^3$ of the coupling constant inside $R$.
On its own this norm does not
carry enough information to control the action of the renormalization
group because it only depends on the size of $\phi$ derivatives of $R$
at $\phi =0$.  The renormalization group involves convolution by the
Gaussian measure $\mu_\Gamma $.  The role of the second norm
$\|\cdot\|_{h,G_{\kappa},\mathcal{A}}$ is to control $R$ when it is
tested on the large fields in the tail of $\mu_{\Gamma}$.

A typical polymer amplitude generated by the expansion
of~\cite[Section 3.1]{BMS} (see also our Section \ref{section_expansion})
is of the form $\ph_1\cdots\phi_k e^{-V(\ph)}$
where $\ph_1,\ldots,\ph_k$ refer to the evaluations of the background field
$\ph$ at various locations $x_1,\ldots,x_k$. The latter eventually are
integrated over against a kernel $\cK(x_1,\ldots,x_k)$.
Such $\ph_i$ factors usually need to be estimated pointwise.
This requires a two-step argument (see~\cite[Lemma 5.1]{BMS}).
One bounds the difference between $\ph_i$ and the average of $\ph$
over some polymer using the large field regulator $G_\ka$
which only involves $L^2$ norms of derivatives of $\ph$ but not $\ph$
itself. Then the average value of $\phi$ is controled, via H{\"o}lder's
inequality, thanks to a fraction of the $e^{-g\int \ph^4}$
which is extracted from $e^{-V(\ph)}$ by~\cite[Lemma 5.5]{BMS}.
The cost of the operation is a large $\bar{g}^{-\frac{1}{4}}$ factor
per $\ph_i$.

Note that by the choice of $Q$ in Section \ref{theQ}
the action of the renormalization group keeps
$K=e^{-V}Q$ 
fixed up to a trivial rescaling
of the coupling constant $g$, in the second order in perturbation theory
approximation.
This ensures that the RG map contribution to $R$ is entirely due
to third and higher orders of
perturbation theory.
Now the expansion in~\cite[Section 3.1]{BMS}
typically produces a collection of vertices
$g[(\ph+\ze)^4-\ph^4]$ which involve at least one fluctuation
field $\ze$.
Therefore, in the worst case scenario,
the contribution of such a vertex to a $||\cdot||_h$
norm bound is $\bar{g}\times(\bar{g}^{-\frac{1}{4}})^3=\bar{g}^{\frac{1}{4}}$.
The $R$ activities which correspond to remainders beyond
second order perturbation theory essentially contain at least
three vertices and satisfy a $\bar{g}^{\frac{3}{4}}$ bound.
The last considerations impose the $\bar{g}^2=\bar{g}^{\frac{11}{4}}\times
\bar{g}^{-\frac{3}{4}}$ multiplicative shift
of the $||\cdot||_h$ norm in the definition of the calibrated triple norm
(\ref{calib_def}).
This in turn affects the number $n_0$ of functional derivatives to
be accounted for in the norms. This number has to be at least
equal to $9$ for the needs of~\cite[Lemma 5.15]{BMS}
which transforms a $||R||_h$ decay into a bound on $|R^\sharp|_{h\ast}$,
using a Taylor expansion of the polymer activities
in the field variable around $\phi=0$.

Once the proper definitions for the polymer activity norms
have been made available, the sequence of estimates
in~\cite[Section 5]{BMS} is for the most part reasonably
straightforward. It successively provides bounds for
activities such as $P$ of (\ref{defofP}) and
$(\cS K)^\natural$ of (\ref{defofSKnat})
which are intermediates
on the way to the final RG product $R'$.
Contour integrations are used for conceptual economy when breaking
$R'$ into pieces to be estimated separately.
They are also used for bounds on the $\tilde{\cE}$, $\cE$ defined
in the extraction step where one would otherwise need
more cumbersome estimates on derivatives of polymer activities
with respect to interpolation parameters.

The crucial estimates of~\cite[Section 5]{BMS}
are~\cite[Corollary 5.25]{BMS} and~\cite[Lemma 5.27]{BMS}
which pertain to the linear part of the $R\rightarrow R'$
map, here denoted by $\cLL^{(g,\mu)}(\cdot)$.
There lies the heart of the renormalization problem
in quantum field theory:
the action of the renormalization group has expanding (relevant)
directions. In the present context these are manifested in (\ref{deflinmap})
which contains a sum over $Y$ small satisfying a constraint.
Consider for example the case where $Y$ is a single cube.
Then the constraint amounts to summing over all small cubes contained
in a fixed cube at the next scale, see the same phenomenon discussed
in~\cite{Rivass}.
The renormalization group inevitably has expanding directions because of
the $L^3$ factor resulting from this summation.
In (\ref{deflinmap}) there are two sums and one of them refers
to $Y$ large. 
Typically, for rather intuitive geometrical reasons,
the number of cubes in a polymer strictly decreases when it is coarse
grained to become the smallest covering by cubes on the next scale.
This geometrical effect is exploited in~\cite[Inequality 2.7]{BMS}
followed by a pin and sum argument~\cite[Lemma 5.1]{BY}
to prove that these so called large polymers
are harmless: they are not part of the expanding direction problem.
However this purely
geometrical effect breaks down in the case of small polymers
(see also~\cite[Lemma 11]{AR}).
A compensating good factor $L^{-\frac{7-\ep}{2}}$ then has to be provided
by the scaling behavior of the activity $J$.
The latter corresponds to the $R$-linear part of
what the perturbation expansion produces, when both terms $R^\sharp$
and counterterms $\tilde{F}_R e^{-\tilde{V}}$ are accounted for.
The proper scaling bound on $J$ proceeds by the clever
double Taylor expansion argument of~\cite[Lemma 15]{BDH}
and~\cite[Corollary 5.25]{BMS}.
Roughly, one expands $J$ in the field variable $\ph$ around zero;
then one expands the fields or test functions appearing in the low order
functional derivative terms, with respect to the
space variable $x$.
The normalization conditions~\cite[Equation 4.37]{BMS}
eliminate the low order terms in the bigrading given
by the degree in $\ph$ and the number of spacial derivatives $\partial$.
The surviving terms have enough $L^{-{\frac{3-\ep}{4}}}$ factors
provided by the $\ph$'s and $L^{-1}$ factors given by the $\partial$'s
not only to beat the $L^3$ volume sum but also to leave
an extra $L^{-\frac{1-\ep}{2}}$ which secures the contractivity
of $\cLL^{(g,\mu)}(\cdot)$ for $L$ large, uniformly in $\ep$.

We may now proceed to the statement of the BMS estimates theorem.
Mind the
order of quantifiers which is important.

\begin{theorem}\ \ 
\label{BMSthm}

$\exists \ka_0>0$,
$\exists L_0\in\NN$,

$\forall\ka\in ]0,\ka_0]$,
$\forall\de\in [0,\frac{1}{6}]$,
$\forall\et\in [0,\frac{3}{16}]$,

$\forall A_g\in ]0,\frac{1}{2}]$,
$\forall A_\mu>0$,
$\forall A_R>0$,
$\forall A_{\gb}>0$,

$\exists c_0>0$,
$\forall c\in]0,c_0]$,

$\exists B_g>0$,
$\exists B_{R\cLL}>0$,

$\forall L\in \NN$ such that $L\ge L_0$,

$\exists B_\mu>0$,
$\exists B_{R\xi}>0$,

$\exists \ep_0>0$,

$\forall \ep\in ]0,\ep_0]$,
$\forall {\gb}\in ]0,A_{\gb}\ep]$,

if one uses the notations
\be
D_g\eqdef \left\{
g\in\CC|\ |g-{\gb}|<A_g {\gb}
\right\}\ ,
\ee
\be
D_\mu\eqdef\left\{
\mu\in\CC|\ |\mu|<A_\mu {\gb}^{2-\de}
\right\}\ ,
\ee
\be
D_R\eqdef\left\{
R\in\cB\cB\cS^\CC|\ |||R|||_{\gb}<A_R {\gb}^{\frac{11}{4}-\et}
\right\}\ ;
\ee

then

\begin{enumerate}
\item
The maps $\xi_g$, $\xi_\mu$, $\xi_R$, are well defined and analytic
on the open set $D_g\times D_\mu\times D_R$ with values
in $\CC$, $\CC$, and $\cB\cB\cS^\CC$ respectively.

\item
The map $(g,\mu,R)\mapsto \cLL^{(g,\mu)}(R)$
is well defined and analytic from
$D_g\times D_\mu\times \cB\cB\cS^\CC$ to $\cB\cB\cS^\CC$.
Besides, for any $(g,\mu)\in D_g\times D_\mu$, the map
$R\mapsto \cLL^{(g,\mu)}(R)$ is linear continuous from
$\cB\cB\cS^\CC$ to itself.

\item
The maps $\xi_g$, $\xi_\mu$, $\xi_R$ send
the real cross-section 
\[
(D_g\cap\RR)\times(D_\mu\cap\RR)\times 
(D_R\cap\cB\cB\cS^\RR)
\]
into $\RR$, $\RR$, and $\cB\cB\cS^\RR$
respectively.

\item
The map $(g,\mu,R)\mapsto \cLL^{(g,\mu)}(R)$
sends $(D_g\cap\RR)\times(D_\mu\cap\RR)\times \cB\cB\cS^\RR$
into $\cB\cB\cS^\RR$.

\item
For any $(g,\mu,R)\in D_g\times D_\mu\times D_R$ one has the estimates
\bea
|\xi_g(g,\mu,R)| & \le & B_g {\gb}^{\frac{11}{4}-\et}\ ,\\
|\xi_\mu(g,\mu,R)| & \le & B_\mu {\gb}^{2}\ ,\\
|||\xi_R(g,\mu,R)|||_{\gb} & \le & B_{R\xi} {\gb}^{\frac{11}{4}}\ .
\eea

\item 
For any $(g,\mu,R)\in D_g\times D_\mu\times \cB\cB\cS^\CC$
one has the estimate
\be
|||\cLL^{(g,\mu)}(R)|||_{\gb}\le
B_{R\cLL} L^{-\lp\frac{1-\ep}{2}\rp}
|||R|||_{\gb}\ .
\label{rlinest}
\ee

\end{enumerate}

\label{BMSestimates}
\end{theorem}

\begin{remark}
We suppressed the reference to a calibrator $\gb$
when mentioning the spaces $\cB\cB\cS^\KK$.
This is because the corresponding statements
do not really depend on the choice of one of the equivalent
norms $|||\cdot|||_{\gb}$.
Also note that the notion of analyticity
we used is the standard one in the Banach
space context (see for instance~\cite[Section 2.3]{Berger}).
Finally remember that the $c$ quantity is the one involved in the
relation (\ref{hgbdep}).
\end{remark}

For the proof of the theorem we refer to~\cite[Section 5]{BMS}.
The statements about the maps being well defined
and analytic will follow from the algebraic nature of the formulae
in Section \ref{algebra}, once the estimates are established.
The statements about the map taking real values are obvious
from the formulae in Section \ref{algebra}.
Now for the estimates,
one should say that it
is the $\gb\sim\ep$ special case of Theorem~\ref{BMSestimates}
which is proven in~\cite{BMS}.
This is because the analysis takes place in the vicinity
of the infrared fixed point where one can assume that the $g$
coupling is almost constant equal to $\gbs=\cO(\ep)$.
In other words, the small $\ep$
parameter is attributed two roles at once:
bifurcation parameter and calibrator.
However, by carefully following~\cite[Section 5]{BMS}, one can
see that the arguments still apply if one dissociates
the two functions.
Therefore
all one needs is to go over and
redo the series of Lemmata from~\cite[Section 5]{BMS},
except that one has
to replace
the hypothesis in Equations (5.1-5.3) of~\cite{BMS}
by the new conditions given by the domains $D_g$, $D_\mu$ and
$D_R$, namely,
\bea
|g-{\gb}| & < & A_g {\gb}\ ,\label{gassump}\\
|\mu| & < & A_\mu {\gb}^{2-\de}\ ,\\
|||R|||_{\gb} & < & A_R {\gb}^{\frac{11}{4}-\et}\ ,
\eea
to which one adds
\be
0<\gb\le A_{\gb} \ep\ ,
\ee
knowing that in the end $\ep$ will be taken to be small, after having
fixed $L$.
Then instead of using powers of $\ep$ in the bounds,
one has to use powers of the calibrator $\gb$ instead.
In Lemmata 5.26
and 5.27 of~\cite{BMS}, one has to use bounds in terms of the
norms $||R||_{h,G_\ka,\cA}$ and $|R|_{h_\ast,\cA}$.
Note that for~\cite[Lemma 5.5]{BMS},
one needs $(\Re g)^{\frac{1}{4}}h$ to be small, which
can be achieved be taking $c$ small
provided $\frac{\Re g}{\gb}$ is bounded from above.
This is guaranteed by our assumption (\ref{gassump}).
Rather than~\cite[Lemma 5.5]{BMS}, the reader might
find it more convenient to use instead specializations
of~\cite[Theorem 1]{BDH}. The latter needs the
ratio $\frac{\Im g}{\Re g}$ to be bounded,
which again is guaranteed by (\ref{gassump}) and the condition
$A_g\le \frac{1}{2}$.
Note that the important~\cite[Equation 5.58]{BMS}
on the other hand cannot allow $(\Re g)^{\frac{1}{4}}h$
to be too small either. This
is why it seems hard to avoid the fibered norm problem,
and we need to keep $g$ rather close to the calibrator $\gb$
as in (\ref{gassump}).
Note that a stronger hypothesis was used in~\cite[Equation 5.1]{BMS}.
However, as far as~\cite[Section 5]{BMS} alone is concerned,
this hypothesis only serves to show that it reproduces
itself, in~\cite[Corollary 5.18]{BMS}. We relaxed this conclusion
in Theorem~\ref{BMSestimates}, and therefore we can
drop this hypothesis.

Remark that in~\cite[Section 5]{BMS} the exponents
$\de,\et$ were taken equal to $\frac{1}{64}$.
The reader who prefers this choice, can simply make
the corresponding modifications in our Section \ref{thebigproof}.
The ranges $[0,\frac{1}{6}]$ for $\de$ and $[0,\frac{3}{16}]$ for $\et$
which we have given come from the following considerations.
First note that the hypothesis $\de,\et>0$ in~\cite[Section 5]{BMS}
is only used in order to absorb some constant factors
in the bounds provided in~\cite[Theorem 1]{BMS}. We do not need
this, since we allow the $B$ factors above.
Then note that each time in~\cite[Section 5]{BMS} one has
a bound with a sum of terms with different powers of 
$\ep$, or rather here $\gb$, one has to pick the dominant term
in the $\de,\et\rightarrow 0$ limit. Collecting the inequalities
on $\de,\et$ which ensure that the term picked is indeed dominant,
one can see that $\de\le\frac{1}{6}$ and
$\et\le\frac{3}{16}$ are sufficient for these inequalities to hold.
Finally, the modifications introduced in our Section~\ref{functspace}
for the functional analytic setting, do not affect the bounds.
One may simply mention that~\cite[Lemma 5.15]{BMS} uses
the Taylor formula with integral remainder. Of course one first has
to apply it in the textbook setting of the space we denoted
by $C_\natural^{n_0}({\rm Fld}(X),\KK)$; and only then,
one can use the sharp norm for the differentials and the
$||\cdot||_{C^2(X)}$ norms
for the fields when performing the bounds.

Armed with the previous remarks, the precise statement
of Theorem~\ref{BMSestimates} to aim for, and some patience,
the reader
with expertise on the techniques from~\cite{BY,BDH,BMS}
will have no difficulty adapting the arguments of~\cite[Section 5]{BMS}.


\section{Elementary estimates on the approximate sequence}
\label{elementary}

This section collects the elementary but crucial
estimates on the sequence $(\gb_n)_{n\in\ZZ}$.

\subsection{The discrete step function lemma}

We firstly need some basic bounds on the sequence.

\begin{lemma}\label{basic}
{\bf The step function behaviour}

\noindent 1)
For any nonnegative integer $n$
\be
\gbs\lp 1-(1-\omz)(1+\omz-L^\ep\omz)^n
\rp
\le \gb_n \le
\gbs\lp 1-(1-\omz)(2-L^\ep)^n
\rp\ .
\ee
\noindent 2)
For any nonpositive integer $n$
\be
\gbs\omz L^{\ep n}
\le \gb_n\le
\gbs\omz\lp
\frac{2}{L^\ep+\sqrt{L^{2\ep}-4\omz(L^\ep-1)}}
\rp^{-n}\ .
\ee
\end{lemma}

\begin{remark}
This simply says that, for $n\rightarrow+\infty$,
$\gb_n$ goes exponentially fast to $\gbs$ and that,
for $n\rightarrow-\infty$, $\gb_n$ goes exponentially fast to
$0$, with a transition or `step' in between.
These exponential rates are very weak in the $\ep\rightarrow 0$ limit.
We need as precise estimates on these rates as we can,
to be used
as input for the following analysis.
Indeed, based on these estimates, we will have
to determine the winner
between close competing effects, as one can see in the next subsections.
This is why we included this otherwise
trivial lemma.
\end{remark}

\begin{proof}
On the interval $[\gb_0,\gbs]$ we define
the two functions $f_{+h}$ and $f_{+l}$ 
by
\bea
f_{+h}(x) & \eqdef & f(\gbs)+(x-\gbs) f'(\gbs)\ ,\\
f_{+l}(x) & \eqdef & f(\gb_0)+(x-\gb_0)\times
\frac{f(\gbs)-f(\gb_0)}{\gbs-\gb_0}\ .
\eea
Since $f$ is increasing and concave, one has for any $x\in[\gb_0,\gbs]$
\be
\gb_0\le f_{+l}(x)\le f(x)\le f_{+h}(x)\le \gbs\ .
\ee
A trivial iteration then implies
\be
\begin{aligned}
\forall n\in\NN, \forall x\in[\gb_0,\gbs], & {} \\
& \gb_0\le (f_{+l})^n(x)\le f^n(x)\le (f_{+h})^n(x)\le\gbs\ .
\end{aligned}
\ee
Now note that
\bea
(f_{+h})^n(x)& = &\gbs+(x-\gbs)[f'(\gbs)]^n \\
 & = & \gbs+(x-\gbs)(2-L^\ep)^n\ .
\eea
Likewise
\be
f_{+l}(x)=\gbs+(x-\gbs)\lp
\frac{\gbs-\gb_1}{\gbs-\gb_0}
\rp^n\ .
\ee
Let $\gb_1=\om_1\gbs$, for $\om_1\in]0,1[$,
then
\be
\gb_1=f(\gb_0)=\omz\gbs(L^\ep-L^{2\ep}a\omz\gbs)\ ,
\ee
or
\be
\om_1=\omz(L^\ep-\omz(L^\ep-1))=L^\ep\omz-L^\ep\omz^2+\omz^2\ ,
\ee
so
\bea
\frac{\gbs-\gb_1}{\gbs-\gb_0} & = & \frac{1-\om_1}{1-\omz}\\
 & = & \frac{1-L^\ep\omz+L^\ep\omz^2-\omz^2}{1-\omz}\\
 & = & \frac{(1-\omz)(1+\omz)-L^\ep\omz(1-\omz)}{1-\omz}\\
 & = & 1+\omz-L^\ep\omz\ .
\eea
Thus,
\be
(f_{+l})^n(x)=\gbs+(x-\gbs)
\lp 1+\omz-L^\ep\omz\rp^n\ .
\ee

Now on the interval $[0,\gb_0]$ we also define,
using the inverse $f^{-1}$,
the two functions $f_{-h}$ and $f_{-l}$ 
by
\bea
f_{-h}(x) & \eqdef & x \times \frac{f^{-1}(\gb_0)}{\gb_0}\ ,\\
f_{-l}(x) & \eqdef & x \times (f^{-1})'(0)\ .
\eea
One has for any $x\in[0,\gb_0]$
\be
0\le f_{-l}(x)\le f^{-1}(x)\le f_{-h}(x)\le \gb_0 
\ee
which trivially iterates into
\be
\begin{aligned}
\forall n\in\NN, \forall x\in[0,\gb_0], & {} \\
& 0\le (f_{-l})^n(x)\le (f^{-1})^n(x)\le (f_{-h})^n(x)\le \gb_0\ .
\end{aligned}
\ee
Now
\be
(f_{-l})^n(x)=L^{-\ep n} x
\ee
and
\be
(f_{-h})^n(x)=\lp\frac{\gb_{-1}}{\gb_0}\rp^n x\ .
\ee
Let $\gb_{-1}=\om_{-1}\gbs$ for $\om_{-1}\in]0,1[$.
The latter
is the smallest of the two solutions of the quadratic
equation
\be
L^\ep (\om_{-1}\gbs)-L^{2\ep} a (\om_{-1}\gbs)^2=\omz\gbs\ ,
\ee
i.e.,
\be
(L^\ep-1) \om_{-1}^2-L^\ep \om_{-1}+\omz=0\ ;
\ee
therefore
\be
\om_{-1}  =  \frac{L^\ep-\sqrt{L^{2\ep}-4\omz(L^\ep-1)}}
{2(L^\ep-1)}\ .
\ee
As a result
\bea
(f_{-h})^n(x) & = & \lp\frac{L^\ep-\sqrt{L^{2\ep}-4\omz(L^\ep-1)}}
{2\omz(L^\ep-1)}\rp^n x\\
 & = & \lp
\frac{2}{L^\ep+\sqrt{L^{2\ep}-4\omz(L^\ep-1)}}
\rp^n x\ .
\eea
From the previous considerations, applied to the sequence
$(\gb_n)_{n\in\ZZ}$, the lemma follows. \hfill \end{proof}

This taken care of, we now proceed to the key lemmata
for the construction of a global RG trajectory.

Firstly, the forward `integral equation' (\ref{forwforg})
for $\de g$, or the deviation
of the running coupling constant with respect to the reference sequence
$(\gb_n)_{n\in\ZZ}$, requires an {\em explicit} bound on 
\be
\Si_{\de g-f}(\ep,\ga,\nu)\eqdef
\sup\limits_{n<0}
\left\{
\frac{1}{\gb_n^\ga}
\sum\limits_{n\le p<0}
\gb_p^\nu
\prod\limits_{n\le j\le p} \frac{1}{f'(\gb_j)}
\right\}
\ee
where $\ga,\nu$ are some nonnegative real exponents.

Secondly, the backward `integral equation' (\ref{backforg})
for $\de g$,
requires an analogous bound on
\be
\Si_{\de g-b}(\ep,\ga,\nu)\eqdef
\sup\limits_{n>0}
\left\{
\frac{1}{\gb_n^\ga}
\sum\limits_{0\le p<n}
\gb_p^\nu
\prod\limits_{p<j<n} f'(\gb_j)
\right\}\ .
\ee

Thirdly, the forward `integral equation' (\ref{forwformu})
for $\mu$, or the squared
mass, requires a bound on
\be
\Si_{\mu-f}(\ep,\ga,\nu)\eqdef
\sup\limits_{n\in\ZZ}
\left\{
\frac{1}{\gb_n^\ga}
\sum\limits_{p\ge n}
L^{-\lp\frac{3+\ep}{2}\rp(p-n+1)}
\gb_p^\nu
\right\}\ .
\ee

Fourthly, the backward `integral equation' (\ref{backforR})
for $R$,
or the irrelevant terms generated by the RG transformation,
requires a bound on
\be
\Si_{R-b}(\ep,\ga,\nu)\eqdef
\sup\limits_{n\in\ZZ}
\left\{
\frac{1}{\gb_n^\ga}
\sum\limits_{p<n}
c_R^{n-p-1}
\gb_p^\nu
\right\}
\ee
where $c_R\in]0,1[$ is an upper bound on the operator
norms of the linearized RG maps $\cLL^{(\cdot,\cdot)}$
in the $R$ direction.
We will provide the necessary estimates in reverse order,
i.e., from simple to more involved.

\subsection{The backward bound for $R$}

Assuming the already mentioned hypotheses on $L,\ep, a, c_R, \omz$
we have the following result.

\begin{lemma}
Provided the exponents $\ga,\mu$ satisfy $\nu\ge \ga\ge 0$,
the following inequality holds.
\be
\Si_{R-b}(\ep,\ga,\nu)\le
\BSi_{R-b}(\ep,\ga,\nu)\eqdef
\frac{\gbs^{\nu-\ga}}{1-c_R}\ .
\ee
\end{lemma}

\begin{proof}
Let $n\in\ZZ$ and denote
\be
\De_n\eqdef
\frac{1}{\gb_n^\ga}
\sum\limits_{p<n}
c_R^{n-p-1}
\gb_p^\nu\ .
\ee
Since the sequence $(\gb_n)_{n\in\ZZ}$ contained in $]0,\gbs[$
is increasing, and $\nu\ge\ga\ge 0$, we trivially have
\bea
\De_n & \le & \frac{1}{\gb_n^\ga}
\sum\limits_{p<n}
c_R^{n-p-1}
\gb_n^\nu\\
 & \le & \frac{\gb_n^{\nu-\ga}}{1-c_R}\\
 & \le & \frac{\gbs^{\nu-\ga}}{1-c_R}\ .
\eea
\hfill \end{proof}

\subsection{The forward bound for $\mu$}

Again with the assumptions of Section \ref{sectBBSS}, we have the following
result.
\begin{lemma}
Provided the exponents $\ga,\nu$ satisfy
$\nu\ge \ga\ge 0$, and $\ep\nu<\frac{3+\ep}{2}$, we have
\be
\Si_{\mu-f}(\ep,\ga,\nu)\le
\BSi_{\mu-f}(\ep,\ga,\nu)\eqdef
\frac{\gbs^{\nu-\ga}}{L^{\frac{3+\ep}{2}}-L^{\ep\nu}}\ .
\ee
\end{lemma}
\begin{proof}
Let $n\in\ZZ$ and write
\bea
\De_n & \eqdef &
\frac{1}{\gb_n^\ga}
\sum\limits_{p\ge n}
L^{-\lp\frac{3+\ep}{2}\rp (p-n+1)}
\gb_p^\nu\\
 & = & \gb_n^{\nu-\ga}
\sum\limits_{p\ge n}
L^{-\lp\frac{3+\ep}{2}\rp (p-n+1)}
\lp
\prod\limits_{n<j\le p} \frac{\gb_j}{\gb_{j-1}}
\rp^\nu\ .
\eea
Now
\be
\frac{\gb_j}{\gb_{j-1}}=\frac{f(\gb_{j-1})-f(0)}{\gb_{j-1}-0}=f'(\xi)
>0
\ee
for some $\xi\in]0,\gb_{j-1}[$. Since $f$ is concave
$f'(\xi)\le f'(0)=L^\ep$, and therefore
\bea
\De_n & \le &
\gb_n^{\nu-\ga}
\sum\limits_{p\ge n}
L^{-\lp\frac{3+\ep}{2}\rp (p-n+1)}
L^{\ep\nu(p-n)}\\
 & \le & \gb_n^{\nu-\ga}
L^{-\lp\frac{3+\ep}{2}\rp}
\times
\frac{1}{1-L^{\ep\nu-\lp\frac{3+\ep}{2}\rp}}\ .
\eea
Since $\nu-\ga\ge 0$, $\gb_n^{\nu-\ga}\le \gbs^{\nu-\ga}$, and we are done.
\hfill \end{proof}

\subsection{The backward bound for $\de g$}

Again with the assumptions of Section \ref{sectBBSS},
we have the following
result.
\begin{lemma}
For any $\ga,\nu\ge 0$ we have
\be
\Si_{\de g-b}(\ep,\ga,\nu)\le
\BSi_{\de g-b}(\ep,\ga,\nu)
\ee
where
\be
\BSi_{\de g-b}(\ep,\ga,\nu)\eqdef
\frac{\omz^{-\ga}\gbs^{\nu-\ga}}{L^\ep-1}
\exp\left[
\frac{2(1-\omz)(1+\omz-L^\ep \omz)}{\omz (2-L^\ep)}
\right]\ .
\ee
\end{lemma}

\begin{proof}
Let $n$ be a strictly positive integer, and denote
\be
\De_n\eqdef
\frac{1}{\gb_n^\ga}
\sum\limits_{0\le p<n}
\gb_p^\nu
\prod\limits_{p<j<n} f'(\gb_j)\ .
\ee
Lemma \ref{basic} shows that $\gb_n\rightarrow\gbs$ when $n\rightarrow+\infty$.
We therefore expect most of the $f'(\gb_j)$ to be very close to
$f'(\gbs)=2-L^\ep$. This motivates the rewriting
\be
\De_n=\frac{1}{\gb_n^\ga}
\sum\limits_{0\le p<n}
\left\{
\prod\limits_{p<j<n} \frac{f'(\gb_j)}{2-L^\ep}
\right\}
(2-L^\ep)^{n-p-1}
\gb_p^\nu\ .
\ee
Since $f'$ is decreasing, for any $j\ge 1$
\be
\frac{f'(\gb_j)}{2-L^\ep}=\frac{L^\ep-2 L^{2\ep} a \gb_j}{2-L^\ep}>1\ ,
\ee
and thus
\bea
\prod\limits_{p<j<n} \frac{f'(\gb_j)}{2-L^\ep} & \le & \prod\limits_{j\ge 1}
\frac{L^\ep-2 L^{2\ep} a \gb_j}{2-L^\ep} \\
 & \le & \exp\left[
\sum\limits_{j\ge 1} \lp
\frac{L^\ep-2 L^{2\ep} a \gb_j}{2-L^\ep} -1
\rp
\right]\ .
\eea
Now
\be
\frac{L^\ep-2 L^{2\ep} a \gb_j}{2-L^\ep} -1=
\frac{2L^{2\ep}a}{2-L^\ep}
\times(\gbs-\gb_j)
\ee
and Lemma \ref{basic} implies
\be
\gb_j\ge \gbs-\gbs(1-\omz)(1+\omz-L^\ep \omz)^j\ ,
\ee
i.e.,
\be
\frac{L^\ep-2 L^{2\ep} \gb_j}{2-L^\ep} -1
\le
\frac{2L^{2\ep}a}{2-L^\ep}
\times
\gbs (1-\omz)(1+\omz-L^\ep \omz)^j
\ee
where $1+\omz-L^\ep \omz$ belongs to $]0,1[$.
Hence
\bea
\prod\limits_{p<j<n} \frac{f'(\gb_j)}{2-L^\ep} & \le & \exp
\left[
\frac{2L^{2\ep}a \gbs(1-\omz)}{2-L^\ep}
\times
\frac{(1+\omz-L^\ep \omz)}{1-(1+\omz-L^\ep \omz)}
\right] \nonumber\\
 & & \\
 & \le & \exp\left[
\frac{2(1-\omz)(1+\omz-L^\ep \omz)}{\omz (2-L^\ep)}
\right]\ .
\label{prodgb}
\eea
So we are left with bounding
\be
\De'_n\eqdef
\frac{1}{\gb_n^\ga}
\sum\limits_{0\le p<n}
(2-L^\ep)^{n-p-1} \gb_p^\nu\ .
\ee
To this effect we use the very coarse estimates
$\gb_n\ge \gb_0=\omz\gbs$ and $\gb_p\le \gbs$ with the result that
\bea
\De'_n & \le & (\omz\gbs)^{-\ga}
\sum\limits_{0\le p<n}
(2-L^\ep)^{n-p-1} \gbs^\nu \\
 & \le & \omz^{-\ga}\gbs^{\nu-\ga}\times \frac{1}{1-(2-L^\ep)}\ .
\label{sumgb}
\eea
Inequalities (\ref{prodgb}) and (\ref{sumgb}) now imply
\be
\De_n\le
\frac{\omz^{-\ga}\gbs^{\nu-\ga}}{L^\ep-1}
\exp\left[
\frac{2(1-\omz)(1+\omz-L^\ep \omz)}{\omz (2-L^\ep)}
\right]\ .
\ee
\hfill \end{proof}

\subsection{The forward bound for $\de g$}

Once more,
with the assumptions of Section \ref{sectBBSS}, we have the following
result.
\begin{lemma}\label{forwardg}
For any exponents $\ga,\nu$ such that
$0\le \ga\le 1$, $\nu> 0$
and
\be
\Ups\eqdef
\frac{2 L^{\frac{\ep}{\nu}}}{L^\ep+\sqrt{L^{2\ep}-4\omz(L^\ep-1)}}
\in ]0,1[\ ,
\label{gammacond}
\ee
we have
\be
\Si_{\de g-f}(\ep,\ga,\nu)\le
\BSi_{\de g-f}(\ep,\ga,\nu)
\ee
where
\[
\BSi_{\de g-f}(\ep,\ga,\nu)\eqdef
\frac{(\omz\gbs)^{\nu-\ga}}{1-\Ups^\nu}
\times
\exp\left[
\frac{\omz\lp
2-L^\ep+\sqrt{L^{2\ep}-4\omz(L^\ep-1)}
\rp}{(1-\omz)\lp
L^\ep-2\omz(L^\ep-1)
\rp}
\right]\ .
\]
\be
{\ }
\ee
\end{lemma}

\begin{proof}
Let $n$ be a strictly negative integer, and define
\be
\De_n\eqdef
\frac{1}{\gb_n^\ga}
\sum\limits_{n\le p<0}
\gb_p^\nu
\prod\limits_{n\le j\le p} \frac{1}{f'(\gb_j)}\ .
\ee
Lemma \ref{basic} shows that $\gb_n\rightarrow 0$ when $n\rightarrow-\infty$.
We therefore expect most of the $f'(\gb_j)$ to be very close to
$f'(0)=L^\ep$.
Therefore write
\be
\De_n=\frac{1}{\gb_n^\ga}
\sum\limits_{n\le p<0}
\lp\prod\limits_{n\le j\le p} \frac{L^\ep}{f'(\gb_j)}\rp
\lp L^{-\ep}\rp^{p-n+1}
\gb_p^\nu\ .
\ee
Now
\be
\frac{L^\ep}{f'(\gb_j)}=\frac{1}{1-2L^\ep a \gb_j}>1\ .
\ee
We use
\bea
\prod\limits_{n\le j\le p}
\frac{L^\ep}{f'(\gb_j)} & \le & \prod\limits_{j\le - 1}
\frac{1}{1-2L^\ep a \gb_j} \\
 & \le & \exp\left[
\sum\limits_{j\le -1} 
\lp\frac{1}{1-2L^\ep a \gb_j}-1\rp
\right]\\
 & \le & \exp\left[
\sum\limits_{j\le -1} 
\frac{2L^\ep a \gb_j}{1-2L^\ep a \gb_j}
\right]\ .
\eea
Now for $j\le -1$, $\gb_j\le \gb_0=\omz\gbs$; hence
\be
\frac{2L^\ep a \gb_j}{1-2L^\ep a \gb_j}
\le
\frac{2L^\ep a \gb_j}{1-2L^\ep a \gb_0}
=
\frac{2L^{2\ep} a \gb_j}{L^\ep-2\omz(L^\ep-1)}\ ,
\ee
and by Lemma \ref{basic}
\be
\frac{2L^\ep a \gb_j}{1-2L^\ep a \gb_j}
\le
\frac{2L^{2\ep} a \omz\gbs}{L^\ep-2\omz(L^\ep-1)}
\lp
\frac{2}{L^\ep+\sqrt{L^{2\ep}-4\omz(L^\ep-1)}}
\rp^{-j}\ .
\ee
As a result
\be
\begin{aligned}
{} & \prod\limits_{n\le j\le p} \frac{L^\ep}{f'(\gb_j)} \\
& \le
\exp\left[
\frac{2\omz(L^\ep-1)}{L^\ep-2\omz(L^\ep-1)}\times
\frac{\lp
\frac{2}{L^\ep+\sqrt{L^{2\ep}-4\omz(L^\ep-1)}}
\rp}{1-\lp
\frac{2}{L^\ep+\sqrt{L^{2\ep}-4\omz(L^\ep-1)}}
\rp}
\right]\ .
\end{aligned}
\ee
Note that
\be
0<\frac{2}{L^\ep+\sqrt{L^{2\ep}-4\omz(L^\ep-1)}}<1\ ,
\ee
because of the global assumptions $1<L^\ep<2$ and $0<\omz<1$.
A straightforward simplification of the argument of the exponential
leads to
\be
\prod\limits_{n\le j\le p} \frac{L^\ep}{f'(\gb_j)} \le
\exp\left[
\frac{\omz\lp
2-L^\ep+\sqrt{L^{2\ep}-4\omz(L^\ep-1)}
\rp}{(1-\omz)\lp
L^\ep-2\omz(L^\ep-1)
\rp}
\right]\ .
\label{prodgf}
\ee
Now we are left with bounding
\be
\De'_n\eqdef
\frac{1}{\gb_n^\ga}
\sum\limits_{n\le p<0}
\lp L^{-\ep}\rp^{p-n+1} \gb_p^\nu\ .
\ee
We now use Lemma \ref{basic} to obtain
\be
\begin{aligned}
{} & \De'_n\le (\omz\gbs)^{-\ga} L^{-\ga\ep n}\\
& \times
\sum\limits_{n\le p<0}
\lp L^{-\ep}\rp^{p-n+1}
(\omz\gbs)^{\nu}
\lp
\frac{2}{L^\ep+\sqrt{L^{2\ep}-4\omz(L^\ep-1)}}
\rp^{-\nu p}\ ,
\end{aligned}
\ee
i.e.,
\be
\De'_n\le (\omz\gbs)^{\nu-\ga}
L^{\ep(\ga-1)|n|}\times L^{-\ep}\times
\frac{\Ups^\nu}{1-\Ups^\nu}
\label{sumgf}
\ee
where
$\Ups$ is the one defined in the statement of the lemma.
We now need a bound which is $n$-independent; this requires
the hypothesis $\ga\le 1$.
Inequalities (\ref{prodgf}) and (\ref{sumgf}) now clearly imply
\be
\forall n\le -1,\ \ 
\De_n\le
\BSi_{\de g-f}(\ep,\ga,\nu)\ ,
\ee
and the lemma is proved.
\hfill \end{proof}

\subsection{The $\ep\rightarrow 0$ limit}

Leaving $L,c_R, \omz$ and the exponents $\ga,\nu$ fixed,
we now analize the $\ep\rightarrow 0$ asymptotics
of the previous bounds.
Note that in this limit we will have
$a=a(L,\ep)\rightarrow\frac{\log L}{18\pi^2}$.
The crux of our construction lies in the following
result.
\begin{lemma}
\label{crucial}
For $\ep\rightarrow 0^+$ we have

\noindent 1)
\be
\BSi_{R-b}(\ep,\ga,\nu)=
\ep^{\nu-\ga}
\lp K_{R-b}+o(\ep)\rp
\ee
where
\be
K_{R-b}=\frac{1}{1-c_R}
\lp
18\pi^2
\rp^{\nu-\ga}\ ,
\ee
provided $\nu\ge\ga\ge 0$;

\noindent 2)
\be
\BSi_{\mu-f}(\ep,\ga,\nu)=
\ep^{\nu-\ga}
\lp K_{\mu-f}+o(\ep)\rp
\ee
where
\be
K_{\mu-f}=\frac{1}{L^{\frac{3}{2}}-1}
\lp
18\pi^2
\rp^{\nu-\ga}\ ,
\ee
provided $\nu\ge\ga\ge 0$;

\noindent 3)
\be
\BSi_{\de g-b}(\ep,\ga,\nu)=
\ep^{\nu-\ga-1}
\lp K_{\de g-b}+o(\ep)\rp
\ee
where
\be
K_{\de g-b}=
\frac{\lp 18\pi^2\rp^{\nu-\ga}}{\omz^\ga(\log L)}
\exp\left[
\frac{2(1-\omz)}{\omz}
\right]\ ,
\ee
provided $\nu\ge 0$ and $\ga\ge 0$;

\noindent 4)
\be
\BSi_{\de g-f}(\ep,1,\nu)=
\ep^{\nu-2}
\lp K_{\de g-f}+o(\ep)\rp
\ee
where
\be
K_{\de g-f}=
\frac{\omz^{\nu-1}{\lp 18\pi^2 \rp}^{\nu-1}}{(\log L)\left[
\nu(1-\omz)-1\right]}
\exp\left[
\frac{2\omz}{1-\omz}
\right]\ ,
\ee
provided $\nu>\frac{1}{1-\omz}$.
\end{lemma}

\begin{proof}
Straightforward first year calculus;
the only delicate point is in checking condition
(\ref{gammacond}).
Simply note the asymptotics
\be
\Ups=1-\lp 1-\omz-\frac{1}{\nu}\rp\ep\log L+ o(\ep)\ ,
\ee
in order to check that Lemma \ref{forwardg} applies,
with the above hypothesis on $\nu$. 
\hfill \end{proof}


\section{Fixed point in the space of sequences}
\label{thebigproof}

We start by applying Theorem \ref{BMSestimates}.
So we choose some $\ka_0>0$ and $L_0\in\NN$
whose existence is guaranteed
by the theorem. We set $\ka=\ka_0$,
and we take
\bea
A_g & = & \frac{1}{2}\ ,\\
A_\mu & = & 1\ ,\\
A_R & = & 1\ ,\\
A_{\gb} & = & 19\pi^2\ , \label{agbchoice}\\
\de & = & \frac{1}{6}\ ,\\
\et & = & \frac{3}{16}\ .
\eea
Now take $c$ to be equal to a $c_0$ provided by the theorem,
which also produces some $B_g$ and $B_{R\cLL}$
only depending on the quantities which have been fixed so far.
Now choose $L\ge L_0$ large enough
so that
\be
B_{R\cLL} L^{-\frac{1}{4}}\le \frac{1}{3}\ .
\ee
This will guarantee that for any $\ep\in]0,\frac{1}{2}]$,
\be
B_{R\cLL} L^{-\lp\frac{1-\ep}{2}\rp}\le \frac{1}{3}\ .
\ee
Now the theorem provides us with $B_\mu$, $B_{R\xi}$, and
$\ep_0$.
We will choose some $\ep_1$ such that
$0<\ep_1<\min(\frac{1}{2},\ep_0)$, and such that for all
$\ep\in]0,\ep_1]$ one has $\frac{\gbs}{\ep}<A_{\gb}$.
This is possible thanks to (\ref{gbsequiv}) and (\ref{agbchoice}).

We now have the following specialization of Theorem
\ref{BMSestimates}.

\begin{proposition}\label{specializ}
There exists an $\ep_2\in]0,\ep_1]$ such that
for any $\ep\in]0,\ep_2]$,
and for any calibrator $\gb\in]0,\gbs[$,
the conclusions (1)--(6) of Theorem \ref{BMSestimates}
are valid with the inequality in (\ref{rlinest})
replaced by
\be
|||\cLL^{(g,\mu)}(R)|||_{f(\gb)}\le
\frac{1}{2}
|||R|||_{\gb}\ .
\label{rlinstream}
\ee
\end{proposition}

The proof is an immediate corollary of the following lemma.

\begin{lemma}
Provided
\be
\max\lp
L^{2\ep},(2-L^\ep)^{-\frac{1}{4}}
\rp\le \frac{3}{2}
\ee
which will hold true when $\ep\rightarrow 0$,
one has for any $\gb\in]0,\gbs[$, and
any $R\in\cB\cB\cS^\KK$,
\be
|||R|||_{f(\gb)}\le \frac{3}{2} |||R|||_{\gb}\ .
\label{normcompare}
\ee
\end{lemma}

\begin{proof}
Let $\gb'=f(\gb)$. Since $\gb'>\gb$, and from the definition
of the triple norms it is immediate that for any $R$ one
has
\be
|||R|||_{\gb'}\le\max
\left[
\lp\frac{\gb'}{\gb}\rp^2,
\lp\frac{\gb'}{\gb}\rp^{\frac{7}{4}},
\ldots, 1,
\lp\frac{\gb'}{\gb}\rp^{-\frac{1}{4}}
\right]
\times|||R|||_{\gb}\ .
\ee
However, by the mean value theorem,
\be
\frac{\gb'}{\gb}=\frac{f(\gb)-f(0)}{\gb-0}=f'(\varsigma)
\ee
for some $\varsigma\in]0,\gbs[$.
As a result
\be
2-L^\ep<\frac{\gb'}{\gb}<L^\ep\ ,
\ee
and the Lemma follows.
\hfill \end{proof}

Now given $\omz\in]0,\frac{1}{2}[$, we construct the
sequence $(\gb_n)_{n\in\ZZ}$ as in Section \ref{sectBBSS}, as well
as the associated spaces $(\cB\cB\cS\cS^\KK, ||||\cdot||||)$.
Given an element $\de s\in\cB\cB\cS\cS^\KK$, and a positive
number $\beta$ we use the notation
$B_\KK(\de s,\beta)$
for the open ball of radius $\beta$ around $\de s$ in
$\cB\cB\cS\cS^\KK$. We also use $\bar{B}_\KK(\de s,\beta)$
for the analogous closed ball.
We can now state our main theorem.
\begin{theorem}\label{mainthm}
{\bf The main theorem}

$\exists \beta_0$,
$\forall \beta\in]0,\beta_0]$,

$\exists \ep_3>0$,
$\forall\ep\in]0,\ep_3]$,

one has

\begin{enumerate}
\item
The $\cB\cB\cS\cS^\CC$ valued map $\mF$ from
Section \ref{sectBBSS} is well defined and analytic
on $B_\CC(0,\beta)$.
\item
The image by $\mF$ of $B_\CC(0,\beta)$
is contained in $\bar{B}_\CC(0,\frac{\beta}{6})$.
\item
The restriction of $\mF$ to the closed ball
$\bar{B}_\RR(0,\frac{\beta}{6})$
is a contraction from that ball to itself.
\item
There exists a unique fixed point for the map
$\mF$ inside the ball $\bar{B}_\RR(0,\frac{\beta}{6})$.
\end{enumerate}

\end{theorem}

\begin{proof}
Let $\beta>0$ be such that the
condition $\beta\le \frac{1}{2}=A_g$ is realized.
Then by construction, for any $n\in\ZZ$,
$\gb_n\in]0, A_{\gb} \ep[$.
Therefore, as a consequence of Proposition \ref{specializ},
for any
\[
\de s=(\de g_n,\mu_n,R_n)_{n\in\ZZ}\in
B_\CC(0,\beta)\ ,
\]
all the summands in (\ref{backforg}),
(\ref{forwforg}), (\ref{forwformu}), and
(\ref{backforR}) are well defined and analytic
with respect to $\de s$.
The analyticity property required in statement {\it (1)} will
therefore follow from the uniform absolute convergence
of the series. The latter will in turn result from
the estimates, required for the statement {\it (2)},
which we now proceed to establish.
Using the notations of Definition \ref{defbigmap}, we
assume that $\de s$ is in $B_\CC(0,\beta)$, and we apply the
estimates of Section \ref{elementary}, in order to obtain
the following results.

\medskip
\noindent{\bf The backward {\boldmath $\delta g$} bound :}

Let $n>0$, then
\bea
\lefteqn{
\frac{1}{\beta}|\de g'_n|{\gb_n}^{-\frac{3}{2}}\le
\frac{1}{\beta{\gb_n}^{\frac{3}{2}}}\sum\limits_{0\le p<n}
\lp
\prod\limits_{p<j<n} f'(\gb_j)
\rp
} & & \nonumber\\
 & & \times
\left[
L^{2\ep}a(L,\ep)|\de g_p|^2+
|\xi_{g}(\gb_p+\de g_p,\mu_p,R_p)|
\right] \nonumber\\
 & & \le \frac{1}{\beta{\gb_n}^{\frac{3}{2}}}\sum\limits_{0\le p<n}
\lp
\prod\limits_{p<j<n} f'(\gb_j)
\rp\\
& & \times
\left[
L^{2\ep}a(L,\ep)\beta^2 {\gb}_p^3+
B_g {\gb}_p^{\lp\frac{11}{4}-\frac{3}{16}\rp}
\right]\\
 & & \le
\beta L^{2\ep}a(L,\ep)
\BSi_{\de g-b}\lp\ep,\frac{3}{2},3\rp
+\frac{1}{\beta}
B_g \BSi_{\de g-b}\lp\ep,\frac{3}{2},\frac{11}{4}-\frac{3}{16}\rp\ .
\nonumber \\
 & & \ 
\eea
Now by part {\it 3)} of Lemma \ref{crucial} and for any fixed
$\beta$,
the last upper bound
goes to zero when $\ep\rightarrow 0$.
Therefore, by choosing $\ep$ small enough,
one will have
\be
\forall n>0\ ,\ \frac{1}{\beta}|\de g'_n|
{\gb_n}^{-\frac{3}{2}}\le\frac{1}{6}\ .
\ee

\medskip
\noindent{\bf The forward {\boldmath $\delta g$} bound :}

Let $n>0$, then in the same vein one will have
\be
\frac{1}{\beta}|\de g'_n|{\gb_n}^{-1}\le
\beta L^{2\ep}a(L,\ep)
\BSi_{\de g-f}(\ep,1,2)
+\frac{1}{\beta}
B_g \BSi_{\de g-b}\lp\ep,1,\frac{11}{4}-\frac{3}{16}\rp\ .
\label{dangerous}
\ee
Now here comes the narrowest passage in the proof.
Provided that $\omz\in]0,\frac{1}{2}[$,
the limiting case of part {\it 4)} in Lemma \ref{crucial}
shows that
\be
L^{2\ep}a(L,\ep)
\BSi_{\de g-f}(\ep,1,2)\rightarrow
\frac{\omz}{1-2\omz}
\exp\left[
\frac{2\omz}{1-\omz}
\right]
\ee
when $\ep\rightarrow 0$.
Therefore we need to take
\be
\beta<
\frac{1-2\omz}{6\omz}\exp\left[
-\frac{2\omz}{1-\omz}
\right]\ .
\ee
Then after $\beta$ is fixed accordingly, the first
term in (\ref{dangerous}) will be strictly less than
$\frac{1}{6}$ in the $\ep\rightarrow 0$ limit
while the second term will go to zero, again
by {\it 4)} of Lemma \ref{crucial}.
We will then have
\be
\forall n<0\ ,\ \frac{1}{\beta}|\de g'_n|
{\gb_n}^{-1}\le\frac{1}{6}\ .
\ee

\medskip
\noindent{\bf The forward {\boldmath $\mu$} bound :}

Let $n\in\ZZ$, then by the same reasoning one will have
\be
\frac{1}{\beta}|\mu'_n|
{\gb_n}^{-(2-\frac{1}{6})}\le
\frac{1}{\beta} B_\mu \BSi_{\mu-f}\lp\ep,2-\frac{1}{6},2\rp
\ee
which will go to zero when $\ep\rightarrow 0$,
as results from {\it case 2)} of Lemma \ref{crucial}.
We will then have
\be
\forall n\in\ZZ\ ,\ \frac{1}{\beta}|\mu'_n|
{\gb_n}^{-(2-\frac{1}{6})}\le\frac{1}{6}\ .
\ee

\medskip
\noindent{\bf The backward {\boldmath $R$} bound :}

Let $n\in\ZZ$, then proceed in the same manner
except that the varying norms require a little care.
We have
\bea
\lefteqn{
\frac{1}{\beta}
|||R'_n|||_{{\gb}_n}\times
{\gb_n}^{-\lp\frac{11}{4}-\frac{3}{16}\rp}
} & & \nonumber\\
 & & \le \frac{1}{\beta{\gb_n}^{\lp\frac{11}{4}-\frac{3}{16}\rp}}
\times
\sum\limits_{p<n}
|||\cLL^{(\gb_{n-1}+\de g_{n-1},\mu_{n-1})}
\circ
\cLL^{(\gb_{n-2}+\de g_{n-2},\mu_{n-2})}
\circ\cdots \nonumber\\
 & & \cdots\circ\cLL^{(\gb_{p+1}+\de g_{p+1},\mu_{p+1})}
\lp\xi_{R}(\gb_p+\de g_p,\mu_p,R_p)\rp|||_{\gb_n}\nonumber\\
 & & \le \frac{1}{\beta{\gb_n}^{\lp\frac{11}{4}-\frac{3}{16}\rp}}
\times
\sum\limits_{p<n}
\lp\frac{1}{2}\rp^{n-p-1}\times\frac{3}{2}
\times B_{R\xi}\times{\gb}_p^{\frac{11}{4}}
\eea
where we repeatedly used the inequality (\ref{rlinstream}),
as well as (\ref{normcompare}), and the $\xi_R$ estimate
in item (5) of Theorem \ref{BMSestimates}.
In sum one has
\be
\frac{1}{\beta}
|||R'_n|||_{{\gb}_n}\times
{\gb_n}^{-\lp\frac{11}{4}-\frac{3}{16}\rp}
\le
\frac{3 B_{R\xi}}{2\beta}\times
\BSi_{R-b}\lp\ep,\frac{11}{4}-\frac{3}{16},\frac{11}{4}\rp
\ee
and this goes to zero when $\ep\rightarrow 0$,
as shown in part {\it 1)} of Lemma \ref{crucial},
with $c_R=\frac{1}{2}$.

\medskip
At this point, statements {\it (1)} and {\it (2)} of the
theorem are
proved.

\medskip
\noindent{\bf The contraction property :}

Let $\de s_1\neq \de s_2$ be two elements
of the open ball $B_\CC(0,\frac{\beta}{6})$.
Let
\be
r=\frac{2\beta}{3||||\de s_1-\de s_2||||}\ .
\ee
Then
\be
||||\de s_1-\de s_2||||\le ||||\de s_1||||
+ ||||\de s_2||||\le\frac{\beta}{3}
\ee
implies that $r\ge 2$.
Therefore, if one defines the contour $\ga$ as the
counterclockwise oriented circle or radius $r$ around the origin
in the complex plane; one has by the Cauchy theorem
\be
\mF(\de s_1)-\mF(\de s_2)=
\frac{1}{2\pi i}
\oint\limits_\ga dz
\lp
\frac{1}{z-1}-\frac{1}{z}
\rp
\mF\lp
\de s_2+z(\de s_1-\de s_2)
\rp\ .
\ee
Now for $z\in\ga$ we have
\bea
||||\de s_2+z(\de s_1-\de s_2)|||| & \le &
||||\de s_2||||+r||||\de s_1-\de s_2|||| \\
 & \le & \frac{\beta}{6}+\frac{2\beta}{3}\\
 & < & \beta\ .
\eea
As a result of the already established statement {\it (2)},
one has
\bea
||||\mF(\de s_1)-\mF(\de s_2)|||| & \le & \frac{1}{r-1}\times
\max\limits_{0\le\theta\le 2\pi}
||||\mF(\de s_2+r e^{i\theta}(\de s_1-\de s_2))||||
\nonumber\\
 & & \\
 & \le & \frac{\beta}{6(r-1)}\\
 & \le & \frac{\beta}{3r}
\eea
because $r\ge 2$.
Inserting the definition of $r$ shows that
\be
||||\mF(\de s_1)-\mF(\de s_2)||||
\le\frac{1}{2}\times
||||\de s_1-\de s_2||||\ ,
\ee
i.e., the contraction property.

\medskip
The real ball stability follows from statements {\it (3)}
and {\it (4)} in
Theorem \ref{BMSestimates}/Proposition \ref{specializ}
and Definition \ref{defbigmap}.
Now statement {\it (3)} is proved, and {\it (4)}
follows from the Banach fixed point theorem.
This concludes the proof of the main theorem.
\hfill \end{proof}

\begin{corollary}
\label{fprecovery}
The constructed two-sided trajectory
$(g_n,\mu_n,R_n)_{n\in\ZZ}$
is the unique such sequence inside the ball
$\bar{B}_\RR (0,\frac{\beta}{6})$
of $\cB\cB\cS\cS^\RR$ which solves the recursion (\ref{recursion}).
One has
\be
\lim\limits_{n\rightarrow -\infty}
(g_n,\mu_n,R_n)=(0,0,0)\ ,
\ee
the trivial Gaussian ultraviolet fixed point, and
\be
\lim\limits_{n\rightarrow +\infty}
(g_n,\mu_n,R_n)=(g_\ast,\mu_\ast,R_\ast)\ ,
\ee
the BMS nontrivial infrared fixed point~\cite{BMS}.
\end{corollary}

\begin{proof}
The proof of the first statement is easy and left to the reader.
Note that the statements concerning the
limits for $R_n$ are topological and
do not depend on a particular choice of a calibrated norm
$|||\cdot|||_{\gb}$.
The last statement follows from the possibility
of making $\beta$ as small as we want, provided
$\ep$ is small enough. Indeed, because of the choice of exponent
$\frac{3}{2}$ in (\ref{gexponent}) at the positive end for $n$,
the convergence of $\gb_n$ to $\gbs$ when $n\rightarrow +\infty$
will ensure that for large positive values of $n$,
$(g_n,\mu_n,R_n)$ will fall within the small
domain around the approximate IR fixed point
where the stable manifold has been constructed,
and where the convergence of all one-sided sequences
which remain bounded in the future, towards the IR fixed point,
has been been established~\cite[Section 6]{BMS}.
\hfill \end{proof}


\section{Suggestions for future work}

The following is a list of problems which are natural
continuations of the present work.

\noindent {\bf 1) }
The continuous connecting orbit between
the two fixed points should be the graph of a function
$g\mapsto(\mu(g), R(g))$ with $g$ in the range
$0<g<g_\ast$. In principle, when considering
one of the sequences
$(g_n,\mu_n,R_n)_{n\in\ZZ}$ we constructed
as a function of $g_0$ only,
this map should correspond to the one giving $\mu_0$
and $R_0$ in terms of $g_0$ (which is here provided
in the range $0<g<\frac{\gbs}{2}$).
One could even say that it is also the map
giving $\mu_n$
and $R_n$ in terms of $g_n$, for any $n$, provided
one could do the proper inversions.
Although we did not yet explore this, it seems likely
that by a more refined analysis, one can construct
the full invariant curve connecting the two fixed points.
This would open the door to the investigation, in a constructive
setting, of the old `reparametrization' renormalization
group~\cite{Stuck,GellMann}.
This has so far remained inaccessible in Bosonic constructive
field theory.
In contrast, a continuous RG for Fermions has been
developed
through work initiated in~\cite{Salmcont} and
completed in~\cite{DisertR}.

\noindent{\bf 2) }
If one could answer the first question, then
the immediate one that follows is: what would be the regularity
of this curve? It seems reasonable to conjecture real analyticity
in the range $0<g<g_\ast$.
An interesting question in this regard raised by K.~Gaw\c{e}dzki,
concerns the $C^\infty$ behavior, or not, of this curve
at $g=0^+$.
A similar question was mentioned in~\cite{GKnonren},
related to a possible explanation of the break down
of the traditional
perturbative argument
ruling out nonrenormalizable theories as
consistent~\cite{Redmond,Symanzik,Parisi}.
To gain insight on this issue, consider
the following simplified flow which mimics the behavior of
the RG map considered here:
\[
\ \ \ 
\left\{
\begin{array}{ccc}
\frac{dg}{dt}=\al g-\beta g^2\ ,\\
\frac{d\mu}{dt}=\ga \mu-\de g^2\ .
\end{array}
\right.
\]
If one eliminates the time variable
then the connecting orbit can be expressed
exactly in terms of an incomplete beta function,
which admits a convergent hypergeometric series
representation near $g=0$.
If one rescales $g$ writing $s=\frac{\beta g}{\al}$ 
and letting $\nu=\frac{\ga}{\al}$ then the smoothness
of the orbit at $0^+$ is reduced to that
of the function
\[
\ \ \ \ \ s\mapsto
s^\nu\frac{\pi(\nu-1)}{\sin[\pi(\nu-1)]}
+\frac{s^2}{\nu-2}\times
{}_2F_1
\left[
\begin{array}{c}
1-\nu\ ,\ 2-\nu\\
3-\nu
\end{array} ; s
\right]
\]
at $s=0$,
when $\nu$ is not an integer.
In this case $C^\infty$ behavior is ruled out.
In our setting $\nu$ is roughly given by
\[
\ \ \ 
\frac{L^{\lp\frac{3+\ep}{2}\rp}-1}{L^{\ep}-1}
\]
which is very large.

\noindent{\bf 3) }
The RG map considered in~\cite{BMS} and also here is
in the so-called `formal infinite volume limit'.
With more work one can probably perform the true
scaling limit of the theory, using an appropriate bare
ansatz as in~\cite{GKnonren} for instance.
One should also try to develop a streamlined rigorous
RG framework for the handling of correlation functions,
including those of more general observables, like composite
operators. An important step in this direction was taken
in~\cite{BKeller}.
One should then prove or disprove the existence
of anomalous scaling dimensions not only for the
field $\ph(x)$, but also for
composite operators. In the hierarchical model there is no
anomalous dimension for the field $\ph(x)$ as shown in~\cite{GKcorrel}.
A similar result for the full model
was recently obtained~\cite{MitterOW,Mitterpcom}, together with a preliminary
perturbative calculation which supports the hypothesis of a nonzero
anomalous dimension of order $\ep$ for the composite field $\ph(x)^2$.
Justifying this last statement by a rigorous nonperturbative proof,
however is a tantalizing open problem.
Finally if one can go as far,
the investigation by analytical means of Wilson's
operator product expansion would make a nice
crowning achievement.

\noindent{\bf 4) }
Orthogonal to the RG approach by Brydges and collaborators,
where one tries
to know as little as possible about the irrelevant terms $R$,
there is also the phase space expansion method~\cite{GlimmJ73}
which has become the trademark of the French school of constructive
field theory~\cite{FMRSGross,FMRS,Rivass} (see also~\cite{Battle}).
In this other approach one, on the contrary, tries
to know as much as possible about the explicit structure
of these terms~\cite{AR}.
We therefore hope to have the future opportunity
of investigating the same model as considered here,
with this alternative approach.
The lessons learnt with the methods of Brydges and collaborators
will be useful in this regard. For instance, in~\cite{AR}
the hypothesis of large $L$ was not used and polymers
were allowed which have large gaps in the vertical direction.
Albeit esthetically pleasing, these features lead to additional
technical difficulties which drive one away from maximal simplicity.
The use of strictly short ranged fluctuation covariances 
introduced in~\cite{MitScop}, exploited in~\cite{BMS} as well as
the present article, and systematically developed in~\cite{BGM,BT},
should allow major simplifications in the multiscale phase
space expansions framework.

\noindent{\bf 5) }
Important new methods for dealing with $\ph^4$-type
lattice models, based on Witten Laplacian
techniques, have been developed recently~\cite{Helffer,Sjostrand,BachSM}.
It would be desirable to extend their reach to the case of critical
theories. Although one should bear in mind that according to RG wisdom,
rather than the weakly convex case
(no $\ph^2$ in the bare potential), it is the double well case
(properly adjusted strictly negative $\ph^2$ coupling) which should entail
a power law behaviour of correlations. The result in the present article
adds a new confirmation to this picture.
Indeed, our trajectory which lies
on the critical manifold essentially
has $\mu=\cO(g^{2-\de})$. Undoing the Wick ordering, this means
that the $\ph^2$ coupling is $\mu-6 C(0) g<0$.

\medskip 

\noindent {{\sc Acknowledgements:} \smaller 
I thank D.~C.~Brydges for giving me this exciting
problem to work on, suggesting fruitful directions
to explore, and spending many hours explaining
the technicalities of the RG formalism he developed
together with his collaborators. They laid the groundwork
for the result presented here.
I thank P.~K.~Mitter for many useful discussions
and in particular for his input concerning the completeness
problem solved in Section \ref{functspace}.
I also thank the anonymous referee for suggesting useful improvements.
This research was initiated during an extended
visit to the University
of British Columbia while I benefited from a relief
from teaching duties or D\'el\'egation during the academic
year 2002/2003 provided by the Centre National de la Recherche
Scientifique.
Most of the work was done during a longer two year visit
to UBC. This was arranged thanks
to the help of D.~C.~Brydges,
J.~Feldman, and G.~Slade.
I thank the UBC Mathematics Department
for providing ideal working conditions.
I also thank my home department the Laboratoire Analyse,
G\'eometrie et Applications of the Universit\'e Paris 13
for giving me leave of absence during that period.}



\vspace{2cm}
\parbox{6cm}{\small 
{\sc Abdelmalek Abdesselam} \\
LAGA, Institut Galil\'ee \\ CNRS UMR 7539\\
Universit{\'e} Paris XIII\\
99 Avenue J.B. Cl{\'e}ment\\
F93430 Villetaneuse \\ France. \\
{\tt abdessel@math.univ-paris13.fr}}

\end{document}